\documentclass[a4paper,12pt]{article}
\pdfoutput=1
\usepackage{graphicx, rotating,amsmath}
\usepackage{hyperref}
\usepackage{slashed}
\usepackage{ifpdf}
\usepackage{cite}
\newcommand{\MS}{\overline{\mbox{\sc ms}}}

\newcommand{\Mhexp}{125.15\pm0.25\GeV}
\newcommand{\Mhnoerr}{125.15\GeV}

\def\eqg#1{eq.~(\ref{#1})}
\newcommand\mpl{M_{\rm Pl}}
\ifx\pdfoutput\undefined
\usepackage[dvips,bookmarks=false]{hyperref}	% This is for arXiv.org
\else
\usepackage{hyperref}	% This is for pdftex
\fi
\hypersetup{colorlinks,bookmarksopen,bookmarksnumbered,citecolor=verdes,
linkcolor=blus,pdfstartview=FitH,urlcolor=rossos}
\def\hhref#1{\href{http://arxiv.org/abs/#1}{#1}} % in bibliography
\allowdisplaybreaks
\def\simlt{\stackrel{<}{{}_\sim}}
\def\simgt{\stackrel{>}{{}_\sim}}
\usepackage{amsmath}
\usepackage{slashed}

\long\def\symbolfootnote[#1]#2{\begingroup%
\def\thefootnote{\fnsymbol{footnote}}\footnote[#1]{#2}\endgroup}

\newcommand{\beq}{\begin{equation}}
\newcommand{\eeq}{\end{equation}}
\newcommand{\fig}[1]{~\ref{fig:#1}}

\newcount\Mac  \Mac=1  % devo mettere Mac=1 se sto lavorando sul file Mac
\newcommand{\ifMac}[2]{\ifnum\Mac=1 #1 \else #2 \fi}
\def\putps(#1,#2)(#3,#4)#5#6{\ifnum\Mac=1 \put(#1,#2){\special{picture #5}}
\else  \put(#3,#4){\includegraphics{#6}} \fi}

\newcommand{\One}{\hbox{1\kern-.24em I}}

\newcommand{\M}{{\cal M}}
\newcommand{\GeV}{\,{\rm GeV}}
\newcommand{\TeV}{\,{\rm TeV}}

\newcommand{\eq}[1]{~{\rm(\ref{eq:#1})}}

\newcommand{\lascia}[1]{}
\makeatletter
\def\art{\@ifnextchar[{\eart}{\oart}}
\def\eart[#1]#2#3#4#5#6{{\rm #2}, {#3 #4} {\rm (#6) #5} [arXiv:{\hhref{#1}}]}
\def\hepart[#1]#2{{\rm #2, arXiv:\hhref{#1}}}
\newcommand{\oart}[5]{{\rm #1}, {#2 #3} {\rm (#5) #4}}

\newcounter{alphaequation}[equation]
\def\thealphaequation{\theequation\hbox to
0.6em{\hfil\alph{alphaequation}\hfil}}
\def\eqnsystem#1{
\def\@eqnnum{{\rm (\thealphaequation)}}
\def\@@eqncr{\let\@tempa\relax \ifcase\@eqcnt \def\@tempa{& & &} \or
  \def\@tempa{& &}\or \def\@tempa{&}\fi\@tempa
  \if@eqnsw\@eqnnum\refstepcounter{alphaequation}\fi
\global\@eqnswtrue\global\@eqcnt=0\cr}
\refstepcounter{equation} \let\@currentlabel\theequation \def\@tempb{#1}
\ifx\@tempb\empty\else\label{#1}\fi
\refstepcounter{alphaequation}
\let\@currentlabel\thealphaequation
\global\@eqnswtrue\global\@eqcnt=0 \tabskip\@centering\let\\=\@eqncr
$$\halign to \displaywidth\bgroup \@eqnsel\hskip\@centering
$\displaystyle\tabskip\z@{##}$&\global\@eqcnt\@ne
\hskip2\arraycolsep\hfil${##}$\hfil& \global\@eqcnt\tw@\hskip2\arraycolsep
$\displaystyle\tabskip\z@{##}$\hfil
\tabskip\@centering&\llap{##}\tabskip\z@\cr}

\def\endeqnsystem{\@@eqncr\egroup$$\global\@ignoretrue} \makeatother

\def\SU{{\rm SU}}

\def\circa#1{\,\raise.3ex\hbox{$#1$\kern-.75em\lower1ex\hbox{$\sim$}}\,}

\usepackage{multicol}
\usepackage{color}
\definecolor{rosso}{cmyk}{0,1,1,0.4}
\definecolor{rossos}{cmyk}{0,1,1,0.55}
\definecolor{rossoc}{cmyk}{0,1,1,0.2}
\definecolor{blu}{cmyk}{1,1,0,0.3}
\definecolor{blus}{cmyk}{1,1,0,0.6}
\definecolor{bluc}{cmyk}{1,1,0,0.1}
\definecolor{verde}{cmyk}{0.92,0,0.59,0.25}
\definecolor{verdec}{cmyk}{0.92,0,0.59,0.15}
\definecolor{verdes}{cmyk}{0.92,0,0.59,0.4}
\definecolor{grigio}{cmyk}{0,0,0,0.07}
\definecolor{rosa}{cmyk}{0,0.1,0.1,0.02}
\definecolor{rosino}{cmyk}{0,0.05,0.05,0.02}
\definecolor{rosas}{cmyk}{0,0.3,0.25,0.05}
\definecolor{celeste}{cmyk}{0.1,0,0,0.02}
\definecolor{giallino}{cmyk}{0,0,0.4,0.02}
\definecolor{rosso}{cmyk}{0,1,1,0.4}
\definecolor{rossos}{cmyk}{0,1,1,0.55}
\definecolor{rossoc}{cmyk}{0,1,1,0.2}
\definecolor{blu}{cmyk}{1,1,0,0.3}
\definecolor{bluc}{cmyk}{1,1,0,0.1}
\definecolor{blucc}{cmyk}{0.7,0.5,0,0}
\definecolor{viola}{cmyk}{0,1,0,0.6}
\definecolor{viola2}{cmyk}{0,1,0.2,0.6}
\definecolor{verde}{cmyk}{0.92,0,0.59,0.25}
\definecolor{verdec}{cmyk}{0.92,0,0.59,0.15}
\definecolor{verdes}{cmyk}{0.92,0,0.59,0.4}
\definecolor{verdino}{cmyk}{0.12,0,0.09,0.05}
\definecolor{giallo}{cmyk}{0,0,1,0}
\definecolor{gialloverde}{cmyk}{0.44,0,0.74,0}

\font\tenrsfs=rsfs10 at 12pt

\font\sevenrsfs=rsfs7
\font\fiversfs=rsfs5
\newfam\rsfsfam
\textfont\rsfsfam=\tenrsfs
\scriptfont\rsfsfam=\sevenrsfs
\scriptscriptfont\rsfsfam=\fiversfs
\def\mathscr#1{{\fam\rsfsfam\relax#1}}

\def\vev#1{\langle {#1} \rangle}

\def\eq#1{eq.~(\ref{#1})}
\def\beq{\begin{equation}}
\def\eeq{\end{equation}}
\def\bea{\begin{eqnarray}}
\def\eea{\end{eqnarray}}
\def\tm{{\tilde m}}
\def\MS{\tm}
\def\M5{m_{1/2}}
\def\Mbino{M_{\widetilde {\scriptscriptstyle B}}}
\def\Mwino{M_{\widetilde {\scriptscriptstyle W}}}
\def\Mgino{M_{\tilde g}}

\newcommand{\sps}{Split SUSY}
\newcommand{\spsh}{Split-SUSY}
\newcommand{\HSS}{High-Scale SUSY}
\newcommand{\msbar}{\overline{\rm MS}}
\newcommand{\drbar}{\overline{\rm DR}}
\newcommand{\xtt}{\widetilde{X}_t}
\newcommand{\xqu}{x_{\scriptscriptstyle QU}}
\newcommand{\smallA}{{\scriptscriptstyle A}}
\newcommand{\smallH}{{\scriptscriptstyle H}}
\newcommand{\mA}{m_\smallA}
\newcommand{\mH}{m_\smallH}
\newcommand{\MH}{M_h}
\newcommand{\sdt}{s_{2\theta}}
\newcommand{\wt}{\widetilde}
\newcommand{\smallGUT}{{\rm {\scriptscriptstyle GUT}}}
\newcommand{\MGUT}{M_\smallGUT}

\def\mt{m_t}
\def\t{\mt^2}

\def\tu{m_{\tilde{t}_1}^2}
\def\td{m_{\tilde{t}_2}^2}
\def\tul{m_{\tilde{t}_1}}
\def\tdl{m_{\tilde{t}_2}}

\def\Q{m_{Q_3}^2}
\def\U{m_{U_3}^2}
\def\SM{{\rm SM}}
\def\split{{\rm split}} 
\def\msplit{m_\split}

\begin{document}

\begin{titlepage}

%\mbox{}
%\vspace*{-1.5cm}

    CERN-PH-TH/2014-131\hfill 
    IFUP-TH/2014

\renewcommand{\thefootnote}{\fnsymbol{footnote}}
\color{black}
\vspace{1.5cm}
\begin{center}
{\LARGE\bf\color{magenta}Higgs Mass and Unnatural Supersymmetry}\\
\bigskip\color{black}\vspace{1cm}{
{\large\bf Emanuele Bagnaschi$^{a,b}$}, 
{\large\bf Gian F.\ Giudice$^{c}$},\\[1mm]
{\large\bf Pietro Slavich$^{a,b}$},
{\large\bf  Alessandro Strumia$^{d,e}$}
} \\[1cm]
{\it (a) LPTHE, UPMC Paris 6, 
  Sorbonne Universit\'es, 4 Place Jussieu, F-75252 Paris, France}\\[1mm]
{\it (b) LPTHE, CNRS, 4 Place Jussieu, F-75252 Paris, France}\\[1mm]
{\it (c) CERN, Theory Division, CH-1211 Geneva 23, Switzerland}\\[1mm]
{\it (d) Dip.\,di Fisica dell'Universit{\`a} di Pisa and INFN, Largo B.\,Pontecorvo 3, I-56127 Pisa, Italy}\\[1mm]
{\it (e) NICPB, Akadeemia tee 23, 12618 Tallinn, Estonia}\\[3mm]
\end{center}
%\symbolfootnote[0]{{\tt e-mail addresses:~~bagnaschi@lpthe.jussieu.fr, 
%    Gian.Giudice@cern.ch,}}
%\symbolfootnote[0]{{\tt  ~~~~~~~~~~~~~~~~~~
%slavich@lpthe.jussieu.fr, astrumia@cern.ch}}
\vskip 0.5cm
\begin{quote}\large
\centerline{\bf Abstract}

\bigskip

  Assuming that supersymmetry exists well above the weak scale, we
  derive the full one-loop matching conditions between the SM and the
  supersymmetric theory, allowing for the possibility of an
  intermediate \spsh\ scale.  We also compute two-loop QCD
  corrections to the matching condition of the Higgs quartic coupling.
  These results are used to improve the calculation of the Higgs mass
  in models with high-scale supersymmetry or split supersymmetry,
  reducing the theoretical uncertainty. We explore the
  phenomenology of a mini-split scenario with gaugino masses
  determined by anomaly mediation. Depending on the value of the
  higgsino mass, the theory predicts a variety of novel possibilities
  for the dark-matter particle.
\end{quote}
\vfill
\end{titlepage}

\setcounter{footnote}{0}
\tableofcontents

\section{Introduction}

The negative results of the searches for new physics at the LHC have
cast some doubts on the existence of low-energy supersymmetry (SUSY) and,
more generally, on the validity of the naturalness principle for the
Fermi scale. However, supersymmetry finds other justifications beyond
naturalness: as a dark matter (DM) candidate, as an element for gauge
coupling unification, as an ingredient for stabilizing the potential
from unwanted vacua at large Higgs field value, or as an ingredient of
superstring theory. This has motivated renewed interest in
``unnatural" setups, in which supersymmetry does not fully cure the
Higgs naturalness problem. In this context, the Higgs mass
measurement~\cite{HiggsMass} has become a crucial (and sometimes the
only) link between theory and experiment. This motivates our detailed
study of the Higgs mass prediction in theories with unnatural
supersymmetry. In particular, we will consider:
\begin{itemize}
\item {\bf Quasi-natural SUSY}, in which supersymmetric
  particles are heavier than the weak scale, but not too far from it
  (say in the $1\!-\!30$~TeV range);

\item {\bf \HSS}, in which all supersymmetric
  particles have masses around a common scale $\tm$, unrelated to
  the weak scale;

\item {\bf \sps}, in which only the scalar
  supersymmetric particles have masses of the order of $\tm$,
  while gauginos and higgsinos are lighter, possibly with
  masses near the weak scale;

\item {\bf Mini-split with anomaly mediation}, in which gauginos get
  mass from anomaly mediation at one loop and scalars from tree-level
  interactions.
\end{itemize} 
Accurate codes have been developed to compute the SUSY
  prediction for the Higgs mass in the natural scenario where $\tilde
  m \approx M_Z$.  When considering the unnatural scenario $\tilde
  m\gg M_Z$, such codes often become redundant and inaccurate: redundant
  because one can ignore effects suppressed by powers of
  $M_Z/\tilde{m}$; inaccurate because one needs to resum large
  logarithms of the ratio $\tm/M_Z$. The computation needs to be
  reorganized: the heavy particles are integrated out at the scale
  $\tm$, where they only induce threshold corrections (free of large
  logarithms) to the SUSY predictions for the couplings of the
  effective theory valid below $\tm$; suitable renormalization-group
  equations (RGEs) are used to evolve the couplings between the
  matching scale $\tm$ and the weak scale, where the running couplings
  are related to physical observables (i.e., the Higgs-boson mass, as
  well as the masses of fermions and of gauge bosons) via Standard
  Model (SM) calculations such as the one in ref.~\cite{SMpar}.

  In this work we improve on the calculation of the threshold
  corrections at the scale $\tm$ by providing complete one-loop
  expressions for all the couplings relevant to the Higgs-mass
  calculation, as well as the dominant two-loop SUSY-QCD corrections
  to the quartic Higgs coupling $\lambda$.

Furthermore, we revisit the tuning condition in the case of \sps, and
we explore mini-split models with anomaly mediation, studying new
possibilities for the LSP, which open new options for the DM
candidate.

\section{Threshold corrections from heavy superparticles}
\label{sec:thre}

In this section we summarize the matching conditions for the couplings
of the effective lagrangian in scenarios where some (if not all) of
the supersymmetric particles are integrated out at the scale $\tm$. We
work under the ``unnatural'' assumption $\tm \gg M_Z$, which induces
significant simplifications with respect to the general expressions
that hold in the natural scenario where $\tm\approx M_Z$.  We complete
and correct the results already presented
in~\cite{Slavich,Giudice:2011cg}.

\subsection{Lagrangian and tree-level matching}
We consider scenarios in which all of the sfermions, as well as a
heavy Higgs doublet $A$, are integrated out at the scale $\tm$. The
surviving (and SM-like) Higgs doublet $H$ is a combination of the two
doublets $H_u$ and $H_d$ of the underlying supersymmetric theory:
\beq
\label{rot}
\left(
\begin{array}{c}
\!\!H\!\!\\\!\!A\!\!
\end{array}\right)~=~
\left(
\begin{array}{cc}
~~\cos\beta&\sin\beta\\
-\sin\beta&\cos\beta
\end{array}\right)
\left(
\begin{array}{c}
\!\!-\epsilon H_d^* \!\\~H_u\!
\end{array}\right)~,
\eeq
where $\epsilon$ is the antisymmetric tensor with $\epsilon_{12}=1$.
The mass parameter $\mA^2$ for the heavy doublet is of the order of
$\tm^2$, whereas the mass parameter $\mH^2$ for the light doublet is
negative and of the order of the weak scale. The potential for the
doublet $H$ below the scale $\tm$ is given by the Standard Model
expression
\beq
V(H) = \frac{\lambda}{2}  \left( H^\dagger H -v^2\right)^2,
\eeq
where $v\approx174$~GeV. The tree-level mass of the physical Higgs
scalar $h$ is $\MH^2=2\lambda v^2$.  The tree-level matching with the
full supersymmetric theory at the scale $\tm$ determines the boundary
condition for the quartic coupling 
\beq
\label{matchlam}
\lambda (\tm )= \frac14\left[g_2^2(\tm)+\frac35\, g_1^{2}(\tm)\right] 
\cos^22\beta~,  
\eeq
where $g_1$ and $g_2$ are the weak gauge coupling constants, assuming
the SU(5) normalization for the hypercharge.  Furthermore, the
tree-level matching condition for the top Yukawa coupling is $g_t(\tm)
= y_t(\tm)\,\sin\beta$, where $y_t$ denotes the coupling of the MSSM
while $g_t$ denotes the coupling of the low-energy effective theory.

We give expressions that can also be applied to the \spsh\ scenario,
where the fermionic superparticles are assumed to be lighter than the
scalars. In such a case the effective lagrangian below the scale $\tm$
includes mass terms for the gauginos and the higgsinos, as well as
Higgs-higgsino-gaugino Yukawa interactions:
\bea
\label{lagr}
{\cal L^{\rm \,split}} &\supset&
-\frac{M_3}2 \tilde g^\smallA\tilde g^\smallA
-\frac{M_2}2 \tilde W^a\tilde W^a
-\frac{M_1}2 \tilde B\tilde B
~-~\mu\, \tilde H_u^T\epsilon \tilde H_d\nonumber +\\
&&- \frac{H^\dagger}{\sqrt{2}}\left( \tilde{g}_{\rm 2u}\sigma^a 
{\tilde W}^a+ \tilde{g}_{\rm 1u}{\tilde B}\right) {\tilde H}_u -
\frac{H^T \epsilon}{\sqrt{2}}\left( -\tilde{g}_{\rm 2d}\sigma^a 
{\tilde W}^a+ \tilde{g}_{\rm 1d}{\tilde B}\right) {\tilde H}_d 
~+~ {\rm h.c.},
\eea
where gauginos and higgsinos are two-component spinors and
$\sigma^a$ are the Pauli matrices. We consider for simplicity the case
of real gaugino and higgsino mass parameters. The tree-level matching
conditions for the \spsh\ couplings at the scale $\tm$ are:
\bea
\tilde{g}_{\rm 2u}(\tm ) = g_2(\tm ) \sin \beta ~,&&~~~~  
\tilde{g}_{\rm 1u}(\tm ) = \sqrt{3/5}\, g_1(\tm ) \sin \beta ~,\nonumber\\
\label{matchgt}
\tilde{g}_{\rm 2d}(\tm ) = g_2(\tm ) \cos \beta ~,&&~~~~ \tilde{g}_{\rm
  1d}(\tm ) = \sqrt{3/5}\, g_1(\tm ) \cos \beta ~.
\eea
Our results for the one-loop matching conditions should be used as
follows:
\begin{itemize}

\item In the High-Scale SUSY scenario, the MSSM is directly matched
  onto the SM at the scale $\tm$, such that the couplings
  $\tilde{g}_{1\rm d},\tilde{g}_{1\rm u}$, $\tilde{g}_{2\rm
    d},\tilde{g}_{2\rm u}$ and $\lambda$ appearing in all one-loop
  threshold corrections can be replaced by their tree-level values of
  \eq{matchgt} and \eq{matchlam}.

\item In the \spsh\ scenario, two different matchings must be applied:
\[
\begin{array}{c}  
\hbox{SM in $\msbar$} 
\\  g_{1,2,3}, g_t, \lambda  
\end{array}
~~~\longleftrightarrow~~~
\begin{array}{c}   
\hbox{\spsh\  in $\msbar$} \\
g_{1,2,3}, g_t, \lambda, \tilde{g}_{1\rm d},\tilde{g}_{1\rm u},
\tilde{g}_{2\rm d},\tilde{g}_{2\rm u}
\end{array}
~~~\stackrel{\tm}{\longleftrightarrow}~~~
 \begin{array}{c}   
\hbox{MSSM in $\drbar$}\\  
g_{1,2,3},y_t 
\end{array}  
\] 
  
The intermediate theory contains higgsinos and gauginos.  Thereby,
their contributions must be removed from the matching conditions at
$\tm$, and included at the lower energy scale at which \sps\ is
matched onto the SM.
\end{itemize}

\subsection{One-loop matching}

To extend our analysis of heavy-SUSY scenarios beyond the leading
order, we need to include in the matching conditions for the couplings
the threshold corrections arising when the heavy particles are
integrated out of the effective low-energy lagrangian. A one-loop
computation of the matching conditions also requires that we specify a
renormalization scheme for the parameters entering the tree-level
part, and include appropriate counterterm contributions in the
one-loop part.

In the full supersymmetric theory above the matching scale $\tm$,
eqs.~(\ref{matchlam}) and (\ref{matchgt}) are valid beyond tree level
only if the parameters are renormalized in a SUSY-preserving
scheme such as $\drbar$.  However, to allow for the direct
implementation of existing SM results in our calculations, we express
all the couplings of the low-energy lagrangian, including the weak
gauge couplings entering the right-hand side of eqs.~(\ref{matchlam})
and (\ref{matchgt}), as running parameters renormalized in the
$\msbar$ scheme. Since this scheme breaks supersymmetry, the
conditions relating the gaugino and four-scalar couplings to the gauge
couplings are not preserved beyond tree level even in the full
supersymmetric theory~\cite{Martin:1993yx}. In the $\msbar$ scheme the
one-loop matching conditions of the gaugino and Higgs-quartic
couplings must therefore be modified as described
in~\cite{Slavich}. In addition, we choose to express the
right-hand-side of eqs.~(\ref{matchlam}) and (\ref{matchgt}) in terms
of the weak gauge couplings of the low-energy theory, as opposed to
the couplings of the full supersymmetric theory. This induces
additional one-loop shifts in the matching conditions in case the
heavy-particle masses are not all equal to $\tm$.

\subsubsection*{Renormalization of $\tan\beta$}

The renormalization of the angle $\beta$ entering
eqs.~(\ref{matchlam}) and (\ref{matchgt}) requires a special
discussion. In contrast to what happens in the MSSM, in the scenarios
considered here it is not useful to relate $\beta$ to the vacuum
expectation values of the Higgs doublets $H_u$ and $H_d$. Instead,
$\beta$ should be interpreted just as a fine-tuned mixing angle that
rotates the two original doublets into a light doublet $H$ and a
massive doublet $A$.
In a generic system of two scalars that mix with each other, the
divergent part of the counterterm for the mixing angle $\theta$ is
fixed by the requirement that it cancel the divergence of the
antisymmetric part of the wave-function renormalization (WFR)
matrix~\cite{mixing}
\beq
\label{mixdiv}
\delta\theta^{\rm \,div} ~=~ \frac12~\frac{\Pi_{12}
^{\rm div}(m_1^2) + \Pi_{12}^{\rm div}(m_2^2)}{m_1^2-m_2^2}~,
\eeq
where $\Pi_{12}^{\rm div}(p^2)$ denotes the divergent part of the
self-energy that mixes the two mass eigenstates characterized by mass
eigenvalues $m_{1,2}^2\,$. The finite part of the counterterm defines
the renormalization scheme for the mixing angle, and different choices
have been discussed in the literature. For example, in
ref.~\cite{mixing} the finite part of the counterterm has the same
form as the divergent part in eq.~(\ref{mixdiv}), while in
ref.~\cite{Espinosa:2001xu} the external momentum in the finite part
of $\Pi_{12}(p^2)$ is set to the special value $(m_1^2+m_2^2)/2$. In
both cases, the renormalized mixing angle $\theta$ is
scale-independent.

In our calculation we define the divergent part of the counterterm
$\delta\beta$ according to eq.~(\ref{mixdiv}), but we choose instead
to define the finite part in such a way that it removes entirely the
contributions of the off-diagonal WFR of the Higgs doublets from the
matching conditions for the effective couplings:
\beq
\label{mixfin}
\delta\beta^{\rm \,fin} ~=~ \frac{\Pi_{\smallH\smallA}^{\rm fin}(\mH^2) 
}{\mH^2-\mA^2}~.
\eeq
Loosely speaking, this defines the renormalized $\beta$ as the angle
that diagonalizes the radiatively corrected Higgs mass matrix at an
external momentum $p^2$ set equal to the light-Higgs mass parameter
$\mH^2$ (in fact, the latter can be considered zero in comparison to
$\mA^2$).

The definition in eq.~(\ref{mixfin}) has the advantage of simplifying
the threshold corrections to the matching conditions, but it leads to
a scale-dependent mixing angle, which at one loop is subject to the
same RGE as the usual parameter $\beta$ of the MSSM. However, it must
be recalled that the angle $\beta$ is not a parameter of the
low-energy lagrangian, and it enters only the matching conditions for
the couplings at the scale $\tm$. Therefore, different choices of
renormalization scheme can be simply compensated for by a shift in the
(arbitrary) input value of $\beta$.
 
\subsubsection*{Threshold corrections to the quartic Higgs coupling} 
 
In the \HSS\ setup where we integrate out all SUSY particles at
the scale $\tm$, the loop-corrected boundary condition for the Higgs
quartic coupling takes the form
\beq
\label{looplam}
\lambda (\tm )= \frac14\left[g_2^2(\tm)+\frac35\, g_1^{2}(\tm)\right] 
\cos^22\beta 
~+~ \Delta \lambda^{1\ell,\,{\rm reg }}
~+~ \Delta \lambda^{1\ell,\,\phi}
~+~ \Delta \lambda^{1\ell,\,\chi^1} 
~+~ \Delta \lambda^{1\ell,\,\chi^2} 
~+~ \Delta \lambda^{2\ell}~,  
\eeq
where we denote by $g_i$ the $\msbar$-renormalized gauge couplings of
the effective theory valid below the scale $\tm$, and $\Delta
\lambda^{1\ell,\,{\rm reg}}$ accounts for the conversion from the
$\overline{\rm DR}$ to the $\overline{\rm MS}$ scheme, which modifies
the tree-level relation of eq.~(\ref{matchlam}) even in the
supersymmetric limit:
\beq
\label{eq:dmsbar}
(4\pi)^2\,\Delta \lambda^{1\ell,\,{\rm reg}} ~=~ 
- \frac{9}{100}\,g_1^4 ~-~ \frac{3}{10} \,g_1^2 g_2^2 
~-\left(\frac{3}{4}-\frac{\cos ^2 2 \beta}{6} \right)  g_2^4 ~.
\eeq
Concerning the other terms in eq.~(\ref{looplam}), $\Delta
\lambda^{1\ell,\,\phi}$ is the one-loop threshold correction arising
when we integrate out the heavy scalars; $\Delta
\lambda^{1\ell,\,\chi^1}$ and $\Delta \lambda^{1\ell,\,\chi^2}$ are
corrections arising when we integrate out the higgsinos and the
electroweak (EW) gauginos; finally, $\Delta \lambda^{2\ell}$ contains
the dominant two-loop correction from diagrams involving stop squarks,
which will be described in the next subsection.

\medskip

Neglecting all Yukawa couplings except the top coupling $g_t$, the
one-loop scalar contribution to the threshold correction to
$\lambda(\tm)$ is\,\footnote{
  As will be explained in section \ref{sec:lambda2loop}, consistency
  with our calculation of the dominant two-loop correction $\Delta
  \lambda^{2\ell}$ requires that the terms of ${\cal O}(g_t^4)$ in
  eq.~(\ref{threshsusy}) be expressed in terms of the
  $\msbar$--renormalized top Yukawa coupling of the low-energy theory
  and of the $\drbar$--renormalized stop masses and mixing.}
\begin{eqnarray}
(4\pi)^2\,\Delta \lambda^{1\ell,\,\phi} &=& 
3 g_t^2 \left[g_t^2 + \frac{1}{2} \left(g_2^2-\frac{g_1^2}5\right) 
\cos  2 \beta  \right] \ln \frac{\Q}{\tm^2} 
   +3 g_t^2  \left[g_t^2 + \frac{2}{5} g_1^2 \cos 2 \beta \right] 
\ln \frac{\U}{\tm^2} \nonumber \\
    &&+\frac{ \cos^2 2 \beta}{300}\, \sum_{i=1}^3\, 
\bigg[3 \left(g_1^4+25 g_2^4\right) 
\ln \frac{m_{Q_i}^2}{\tm^2}  +24 g_1^4 \ln \frac{m_{U_i}^2}{\tm^2} 
+6 g_1^4 \ln \frac{m_{D_i}^2}{\tm^2} \nonumber \\
   && ~~~~~~~~~~~~~~~~~~~~~
+ \left(9 g_1^4+25 g_2^4\right) \ln \frac{m_{L_i}^2}{\tm^2}   
+18 g_1^4 \ln \frac{m_{E_i}^2}{\tm^2}
 \bigg]  \nonumber \\
 &&+\frac{1}{4800}  \bigg[261 g_1^4+630 g_1^2 g_2^2  +1325
 g_2^4  - 4 \cos 4 \beta  \left(9 g_1^4+90 g_1^2 g_2^2+175 g_2^4\right) 
\nonumber \\
&&~~~~~~~~~~~ -9 \cos 8\beta  \left(3 g_1^2+5 g_2^2\right)^2 \bigg] 
\ln \frac{\mA^2}{\tm^2} -\frac3{16} \left(\frac35 g_1^2 + g_2^2\right)^2 
\sin ^2 4 \beta \nonumber \\
   &&+6 g_t^4 \xtt \left[\wt F_1\left(\xqu\right)
     -\frac{\xtt}{12} \wt F_2\left(\xqu\right)\right] 
+ \frac{3}{4} g_t^2 \xtt \cos 2\beta \left[\frac35 g_1^2 
\wt F_3 \left(\xqu\right) 
+ g_2^2 \wt F_4 \left(\xqu\right) \right] \nonumber \\
&&-\frac14  g_t^2 \xtt \cos^2 2\beta 
\left( \frac35 g_1^2 +g_2^2 \right) \wt F_5\left(\xqu \right)
~.
   \label{threshsusy}
\end{eqnarray}
Here: $m_{L_i},\, m_{E_i},\,m_{Q_i},\, m_{U_i}$ and $m_{D_i}$ are the
soft SUSY-breaking masses for the sfermions of the $i$-th generation;
$\xtt \equiv X_t^2/(m_{Q_3}m_{U_3})\,$, where $X_t = A_t-\mu \cot
\beta$ and $A_t$ is the soft SUSY-breaking Higgs-stop coupling; $\xqu
\equiv m_{Q_3}/m_{U_3}\,$; the loop functions $\tilde{F}_i$ are
defined in appendix~\ref{F}, eq.~(\ref{sys:Fstop}).

The first, second and third lines of eq.~(\ref{threshsusy}) contain
threshold corrections arising when the squarks and sleptons are
integrated out of the theory (including the contributions due to the
redefinition of the gauge couplings); the fourth and fifth lines
contain the corresponding contribution of the heavy Higgs doublet; the
last two lines contain the corrections controlled by
$\xtt\,$.\footnote{ Note that the result in eq.~(\ref{threshsusy})
  corrects both eq.~(2.6) of ref.~\cite{Slavich} and eq.~(7) of
  ref.~\cite{Giudice:2011cg}. In the former, a common mass $M_S$ was
  assumed for all of the heavy scalars, therefore most of the terms
  appearing in our eq.~(\ref{threshsusy}) vanish. However, a factor
  $-\!\cos 2\beta$ was omitted in the contribution proportional to
  $h_t^2\,(g^2+g^{\prime\,2})$, and the non-vanishing terms in the
  fifth and in the last lines of our eq.~(\ref{threshsusy}) were
  missed. Concerning eq.~(7) of ref.~\cite{Giudice:2011cg}, the
  heavy-Higgs contribution --- see the fourth and fifth lines of our
  eq.~(\ref{threshsusy}) --- was incorrect, and the term in the last
  line, arising from the diagonal WFR of the external legs, was
  missed. 
  Further discrepancies between our eq.~(\ref{threshsusy}) and
  eq.~(7) of ref.~\cite{Giudice:2011cg} are due to the fact that the
  latter was computed under the assumption that the tree-level part of
  the matching condition is expressed in terms of the
  $\msbar$-renormalized gauge couplings of the MSSM, as opposed to
  those of the low-energy effective theory.
}

Finally, we give the one-loop higgsino-gaugino contributions to the
matching condition for $\lambda(\tm)$.  The first one, containing the
proper threshold corrections to the quartic coupling, was given in
ref.~\cite{Giudice:2011cg} in terms of the effective couplings of
\sps:
\begin{eqnarray}  
\label{eq:hss}
(4\pi)^2\,\Delta \lambda^{1\ell,\,\chi^1} &=&
\frac12 \, \tilde{\beta}_\lambda\, \ln \frac{\mu^2}{\tm^2}    \nonumber
+ \bigg[
-\frac{7}{12}f_1(r_1)   \left(\tilde{g}_{\text{1d}}^4+\tilde{g}_{\text{1u}}^4\right)  
-\frac{9}{4} f_2(r_2) \left(\tilde{g}_{\text{2d}}^4+\tilde{g}_{\text{2u}}^4\right) 
\\  &&  \nonumber
-\frac{3}{2} f_3(r_1) \tilde{g}_{\text{1d}}^2 \tilde{g}_{\text{1u}}^2
-\frac{7}{2} f_4(r_2) \tilde{g}_{\text{2d}}^2   \tilde{g}_{\text{2u}}^2
-\frac{8}{3} f_5(r_1,r_2) \tilde{g}_{\text{1d}} \tilde{g}_{\text{1u}} \tilde{g}_{\text{2d}} \tilde{g}_{\text{2u}}
\\ \nonumber &&
-\frac{7}{6} f_6(r_1,r_2) \left(\tilde{g}_{\text{1d}}^2\tilde{g}_{\text{2d}}^2+\tilde{g}_{\text{1u}}^2 \tilde{g}_{\text{2u}}^2\right)
-\frac{1}{6} f_7(r_1,r_2) \left(\tilde{g}_{\text{1d}}^2
  \tilde{g}_{\text{2u}}^2+\tilde{g}_{\text{1u}}^2 \tilde{g}_{\text{2d}}^2\right)
\\  &&
-\frac{4}{3} f_8(r_1,r_2)
\left(\tilde{g}_{\text{1d}} \tilde{g}_{\text{2u}}+\tilde{g}_{\text{1u}} \tilde{g}_{\text{2d}}\right) \left(\tilde{g}_{\text{1d}}
  \tilde{g}_{\text{2d}}+\tilde{g}_{\text{1u}} \tilde{g}_{\text{2u}}\right)
\nonumber \\   &&   \nonumber
+\frac{2}{3} f\left(r_1\right) \tilde{g}_{\text{1d}}
\tilde{g}_{\text{1u}} \left[\lambda -2 \left(\tilde{g}_{\text{1d}}^2+\tilde{g}_{\text{1u}}^2\right)\right]
+2 f\left(r_2\right) \tilde{g}_{\text{2d}} \tilde{g}_{\text{2u}}
\left[\lambda -2 \left(\tilde{g}_{\text{2d}}^2+\tilde{g}_{\text{2u}}^2\right)\right]
\\  && 
+  \frac{1}{3} g\left(r_1\right) \lambda 
\left(\tilde{g}_{\text{1d}}^2+\tilde{g}_{\text{1u}}^2\right)
+ g\left(r_2\right)  \, \lambda  \left(\tilde{g}_{\text{2d}}^2+\tilde{g}_{\text{2u}}^2\right)   
\bigg]~,
\end{eqnarray}   
where $r_i \equiv M_i/\mu$, and 
\begin{eqnarray}  \nonumber
   \tilde{\beta}_\lambda &=& 2\lambda \left(\tilde{g}_{\text{1d}}^2+\tilde{g}_{\text{1u}}^2+3 \tilde{g}_{\text{2d}}^2+3 \tilde{g}_{\text{2u}}^2\right)
   -\tilde{g}_{\text{1d}}^4
-\tilde{g}_{\text{1u}}^4-5 \tilde{g}_{\text{2d}}^4-5 \tilde{g}_{\text{2u}}^4\\
&&-4 \tilde{g}_{\text{1d}} \tilde{g}_{\text{1u}} \tilde{g}_{\text{2d}} \tilde{g}_{\text{2u}}-2 \left(\tilde{g}_{\text{1d}}^2+\tilde{g}_{\text{2u}}^2\right) \left(\tilde{g}_{\text{1u}}^2+\tilde{g}_{\text{2d}}^2\right)
\end{eqnarray}  
is the \spsh\ contribution to the one-loop beta function of $\lambda$.
The functions $f_i$, $f$ and $g$ are defined in appendix~\ref{F},
eq.~(\ref{sys:f}).
In the case of \HSS, the quartic coupling $\lambda$ and the effective
higgs-higgsino-gaugino couplings entering eq.~(\ref{eq:hss}) must be
expressed in terms of the gauge couplings and of the angle $\beta$ by
means of eqs.~(\ref{matchlam}) and (\ref{matchgt}).

The second higgsino-gaugino contribution to $\lambda(\tm)$,
\beq
(4\pi)^2\,\Delta \lambda^{1\ell,\,\chi^2} ~=~ -   
\frac16\,\cos^22\beta \,\left[
  2\,g_2^4\,\ln\frac{M_2^2}{\tm^2}
  +\left(\frac9{25} g_1^4+g_2^4\right)\ln\frac{\mu^2}{\tm^2}\right]~ ,
\eeq
arises from the fact that in \HSS\ the tree-level part of the matching
condition for $\lambda$ in eq.~(\ref{looplam}) is expressed in terms
of the gauge couplings of the SM.\footnote{This contribution is not
  included in eq.~(7) of ref.~\cite{Giudice:2011cg}, due to the
  different definition adopted in that paper for the gauge couplings
  entering the tree-level matching condition for $\lambda$.}

In the \spsh\ setup, the higgsino-gaugino contributions are removed
from the matching condition for $\lambda(\tm)$, eq.~(\ref{looplam}),
and the tree-level part of the matching condition is expressed in
terms of the gauge couplings of \sps. However, $\Delta
\lambda^{1\ell,\,\chi^1}$ reappears as a threshold correction at the
lower scale $\msplit$ where the \spsh\ lagrangian is matched to the SM
lagrangian:
\beq
\lambda^\SM(\msplit) ~=~  \lambda^\split(\msplit)
~+~\Delta \lambda^{1\ell,\,\chi^1}~.
\eeq

We remark that this procedure neglects effects suppressed by inverse
powers of the superparticle masses, and is therefore accurate only if
there is some hierarchy between the masses of higgsinos and gauginos
and the weak scale. Full one-loop results for the chargino-neutralino
contributions to the Higgs mass in \sps\ were provided in
refs.~\cite{Binger,Slavich}.

We also recall that, in \sps, the soft SUSY-breaking parameter $A_t$
is suppressed by the same symmetry that keeps $\mu$ and the gaugino
masses smaller than the scalar masses. Therefore, the terms
proportional to $\xtt$ in the last two lines of eq.~(\ref{threshsusy})
become negligible.

\subsubsection*{Threshold corrections to the \spsh\ couplings} 
In the \spsh\ scenario one also needs to generalize the tree-level
expressions of eq.~(\ref{matchgt}) for the Higgs-higgsino-gaugino
couplings at the scale $\tm$ adding the one-loop threshold
corrections. We find:
\begin{eqnarray}
\label{eq:g2u}
\frac{\tilde{g}_{\rm 2u}}{g_2\sin\beta} 
= 1 &+& \frac{1}{(4\pi)^2}\bigg\{-g_2^2
\left(\frac{2}{3}+\frac{11}{16}\cos^2\beta\right)
+\frac{3g_1^2}{80}(-2+7\cos^2\beta)+\frac{9\,g_t^2}{4\sin^2\beta} \nonumber \\
&+&\frac{20 g_2^2 
+ 3 (-9 g_1^2 + 35 g_2^2) \cos^2\beta}{120} \ln\frac{\mA^2}{\tm^2} +
\frac{g_2^2}{6} \sum_{i=1}^3 \ln \frac{m_{L_i}^2}{\tm^2} \nonumber \\
&+& \frac{g_2^2}{2} \sum_{i=1}^3 \ln \frac{m_{Q_i}^2}{\tm^2} 
- \frac{3}{4} \frac{g_t^2}{\sin^2\beta} \left[ 3 \ln \frac{m_{Q_3}^2}{\tm^2} - \ln \frac{m_{U_3}^2}{\tm^2} \right]
\bigg\}~,~~\\
\label{eq:g2d}
\frac{\tilde{g}_{\rm 2d}}{g_2\cos\beta} 
= 1 &+& \frac{1}{(4\pi)^2}\bigg\{-g_2^2
\left(\frac{2}{3}+\frac{11}{16}\sin^2\beta\right)
+\frac{3g_1^2}{80}(-2+7\sin^2\beta) 
+ \frac{g_2^2}{2} \sum_{i=1}^3 \ln \frac{m_{Q_i}^2}{\tm^2} \nonumber \\
&+& \frac{20 g_2^2 
+ 3 (-9 g_1^2 + 35 g_2^2) \sin^2\beta}{120} \ln \frac{\mA^2}{\tm^2} +
\frac{g_2^2}{6} \sum_{i=1}^3 \ln \frac{m_{L_i}^2}{\tm^2} 
%\nonumber \\
%&+& \frac{g_2^2}{2} \sum_{i=1}^3 \ln \frac{m_{Q_i}^2}{\tm^2}
\bigg\}~,
\end{eqnarray}
\begin{eqnarray}
\label{eq:g1u}
\frac{\tilde{g}_{\rm 1u}}{\sqrt{3/5}\,g_1\sin\beta} = 
1 &+& \frac{1}{(4\pi)^2}\bigg\{\frac{3g_2^2}{16}(-2+7\cos^2\beta)
+\frac{3g_1^2}{80}(-44+7\cos^2\beta)+\frac{9\,g_t^2}{4\sin^2\beta} \nonumber \\
&+& \frac{4 g_1^2 - 9 (g_1^2 + 5 g_2^2 )\cos^2\beta}{40} \ln \frac{\mA^2}{\tm^2}
+\frac{g_1^2}{10} \sum_{i=1}^3 \left[ \ln \frac{m_{L_i}^2}{\tm^2} + 2 \ln \frac{m_{E_i}^2}{\tm^2} \right]\nonumber \\
&+&\frac{g_1^2}{30}  \sum_{i=1}^3 \left[ \ln \frac{m_{Q_i}^2}{\tm^2} + 8 \ln \frac{m_{U_i}^2}{\tm^2} +
2 \ln \frac{m_{D_i}^2}{\tm^2} \right] \nonumber \\
&+& \frac{g_t^2}{4 \sin^2\beta} \left(7 \ln \frac{m_{Q_3}^2}{\tm^2} - 13 \ln \frac{m_{U_3}^2}{\tm^2} \right)
\bigg\}~,\\\nonumber \\\nonumber \\
\label{eq:g1d}
\frac{\tilde{g}_{\rm 1d}}{\sqrt{3/5}\,g_1\cos\beta} = 
1 &+& \frac{1}{(4\pi)^2}\bigg\{\frac{3g_2^2}{16}(-2+7\sin^2\beta)
+\frac{3g_1^2}{80}(-44+7\sin^2\beta) \nonumber \\
&+& \frac{4 g_1^2 - 9 (g_1^2 + 5 g_2^2) \sin^2\beta}{40} \ln \frac{\mA^2}{\tm^2}
+\frac{g_1^2}{10}  \sum_{i=1}^3 \left[ \ln \frac{m_{L_i}^2}{\tm^2} + 2 \ln \frac{m_{E_i}^2}{\tm^2} \right] \nonumber \\
&+& \frac{g_1^2}{30}  \sum_{i=1}^3 \left[ \ln \frac{m_{Q_i}^2}{\tm^2} + 8 \ln \frac{m_{U_i}^2}{\tm^2} +
2 \ln \frac{m_{D_i}^2}{\tm^2} \right]
\bigg\}~.
\end{eqnarray}

In the equations above we assume that the tree-level part of the
matching conditions is expressed in terms of the $\msbar$-renormalized
couplings of \sps, and that the angle $\beta$ is renormalized
according to the prescription in eq.~(\ref{mixfin}).  
Note that the non-logarithmic terms proportional to $g_t^2$ in
eqs.~(\ref{eq:g2u}) and (\ref{eq:g1u}) and those proportional to
$g_1^2$ and $g_2^2$ in eqs.~(\ref{eq:g1u}) and (\ref{eq:g1d}) differ
from the corresponding terms in eq.~(11) of
ref.~\cite{Giudice:2011cg}, which was based on the results of
ref.~\cite{LG}. The discrepancies can be traced back to the fact that
the renormalization of the angle $\beta$ was neglected in
ref.~\cite{LG}, and to a mistake in eqs.~(B.1) and (B.3) of that
paper.

\subsubsection*{Threshold corrections to the gauge couplings}
Finally, we report the one-loop matching conditions between the
$\msbar$-renormalized gauge and Yukawa couplings of the effective
theory valid below the SUSY scale, $g_{1,2,3}$ and $g_t$, and the
corresponding $\drbar$-renormalized couplings of the MSSM, which we
denote by $\hat g_{1,2,3}$ and $\hat y_t = \hat
g_t/\sin\beta$.\footnote{We can neglect the bottom Yukawa coupling
  because the observed value of the Higgs mass suggests a small
  $\tan\beta$ if the SUSY scale is large, so that $\hat y_b = \hat
  g_b/\cos\beta$ cannot be enhanced by a large $\tan\beta$.}  Such
corrections are not needed for studying the Higgs mass prediction, but
they are needed for studying issues that involve the running couplings
at large energy --- for example gauge coupling unification or the
evolution of the soft parameters above the matching scale $\tm$.

In the \HSS\ scenario, where gauginos and higgsinos are integrated out
at the scale $\tm$ together with the heavy scalars, the threshold
corrections to the gauge couplings are well known:
\bea
\label{dg1}
\hat g_1(\tm) &=&  g_1(\tm) ~ + ~ \frac35\,\frac{g_1^3}{16\pi^2}
\left[ \,
-\frac1{3}\,\ln\frac{\mu^2}{\tm^2}
\,-\,\frac1{12}\,\ln\frac{\mA^2}{\tm^2}
\,-\, \frac{1}{12}\,\sum_{i=1}^3\,\left(
\ln\frac{m^2_{L_i}}{\tm^2}
+ 2\,\ln\frac{m^2_{E_i}}{\tm^2}\right)
\right.\nonumber\\
&&~~~~~~~~~~~~~~~~~~~~~~~~\left. \,-\,
\frac{1}{36}\,\sum_{i=1}^3\,\left(
\ln\frac{m^2_{Q_i}}{\tm^2}
+8\,\ln\frac{m^2_{U_i}}{\tm^2}
+2\,\ln\frac{m^2_{D_i}}{\tm^2}\right)
\right]\,,\\
\nonumber\\
\label{dg2}
\hat g_2(\tm) &=&  g_2(\tm) ~ + ~ \frac{g_2^3}{16\pi^2}\,
\left[\,\frac13 
- \, \frac23\,\ln\frac{M_2^2}{\tm^2}
- \, \frac13\,\ln\frac{\mu^2}{\tm^2}
- \, \frac1{12}\,\ln\frac{\mA^2}{\tm^2}
\right.\nonumber\\
&&~~~~~~~~~~~~~~~~~~~~~~~~~\left.
- \, \frac1{12}\,\sum_{i=1}^3\,\left(
3\,\ln\frac{m^2_{Q_i}}{\tm^2}+ \ln\frac{m^2_{L_i}}{\tm^2}\right)
\right]\,,\\
\nonumber\\
\label{dg3}
\hat g_3(\tm) &=&  g_3(\tm) ~ + ~ \frac{g_3^3}{16\pi^2}\,
\left[\,\frac12 -\, \ln\frac{M_3^2}{\tm^2} 
\,-\, \frac1{12}\,\sum_{i=1}^3\,\left(
2\,\ln\frac{m^2_{Q_i}}{\tm^2}+ \ln\frac{m^2_{U_i}}{\tm^2}
+ \ln\frac{m^2_{D_i}}{\tm^2}\right)\, \right].
\eea
The non-logarithmic terms in eqs.~(\ref{dg2}) and (\ref{dg3}) account
for the $\msbar\,$--$\,\drbar$ conversion of $g_2$ and $g_3$. 

In the \spsh\ scenario the logarithmic terms involving the higgsino
and gaugino masses $\mu$, $M_2$ and $M_3$ must be removed from
eqs.~(\ref{dg1})--(\ref{dg3}), and they reappear at the \spsh\ scale
as threshold corrections between the gauge couplings of \sps\ and the
corresponding couplings of the SM (both defined in the $\msbar$
scheme).

\subsubsection*{Threshold corrections to the top Yukawa coupling}
The one-loop relation between the $\drbar$-renormalized top Yukawa
coupling of the MSSM and the $\msbar$-renormalized coupling of the
effective theory valid below $\tm$ involves a contribution arising
from the $\msbar\,$--$\,\drbar$ conversion, one arising from corrections
involving the heavy scalars, and one arising from corrections
involving only higgsinos and gauginos:
\beq
\label{defdgt}
\hat y_t(\tm) ~=~ \frac{g_t(\tm)}{\sin\beta}\,
\left(1 ~+~ \Delta g_t^{{\rm reg}} ~+~ \Delta g_t^{\phi} 
 ~+~ \Delta g_t^{\chi}\right)~,
\eeq
where the angle $\beta$ entering the tree-level part of the relation
is renormalized according to the prescription in eq.~(\ref{mixfin}).
We find:
\bea
\label{dgtreg}
(4\pi)^2\,\Delta g_t^{{\rm reg}} &=& 
\frac{g_1^2}{120} + \frac{3\,g_2^2}{8} - \frac{4\,g_3^2}{3}~,\\
\nonumber\\
\label{dgtphi}
(4\pi)^2\,\Delta g_t^{\phi}~~ &=& -\,\frac{4}{3}\, g_3^2 \left[
 \ln\frac{M_3^2}{\tm^2} 
+ \wt{F}_6 \left(\frac{m_{Q_3}}{M_3} \right) 
+ \wt{F}_6 \left(\frac{m_{U_3}}{M_3} \right) 
- \frac{X_t}{M_3}\wt{F}_9 
\left(\frac{m_{Q_3}}{M_3},\frac{m_{U_3}}{M_3}\right)
\right] \nonumber \\
&&-\, g_2^2 \,\left[ 
\frac{3}{8} \ln\frac{M_2^2}{\tm^2} 
\,-\, \frac{3}{2} \ln \frac{\mu^2}{\tm^2} 
+\frac{3}{4} \wt{F}_6 \left(\frac{m_{Q_3}}{M_2} \right)
- \frac{3}{4} \wt{F}_8 
\left( \frac{m_{Q_3}}{\mu},\frac{M_2}{\mu} \right) 
\right. \nonumber \\
&&~~~~~~~~~~-\left. \frac{3}{4}\, \frac{M_2}{\mu}\, \cot\beta \,
\wt{F}_9 \left( \frac{m_{Q_3}}{\mu},\frac{M_2}{\mu}\right) 
\right]  \nonumber \\
&&- \frac{3}{5} \,g_1^2 \,\left[\frac{17}{72}\, \ln\frac{M_1^2}{\tm^2} 
\,-\, \frac{1}{2}\, \ln\frac{\mu^2}{\tm^2} 
+ \frac{1}{36}\, \wt{F}_6 \left(\frac{m_{Q_3}}{M_1} \right)
+\frac{4}{9}\, \wt{F}_6 \left( \frac{m_{U_3}}{M_1} \right) 
\right. \nonumber \\
&&  ~~~~~~~~~+\frac{1}{12}\,\wt{F}_8
\left(\frac{m_{Q_3}}{\mu} , \frac{M_1}{\mu}\right) 
-\,\frac13\,\wt{F}_8 \left(\frac{m_{U_3}}{\mu} ,\frac{M_1}{\mu}\right)
-\frac{X_t}{9 M_1}\, 
\wt{F}_9 \left(\frac{m_{Q_3}}{M_1},\frac{m_{U_3}}{M_1} \right) 
\nonumber \\
&&\left. ~~~~~~~~~
+\frac{M_1\,\cot\beta}{12\,\mu}\, 
\wt{F}_9 \left(\frac{m_{Q_3}}{\mu},\frac{M_1}{\mu} \right)
\,-\,\frac{M_1\,\cot\beta}{3 \,\mu}\, 
\wt{F}_9 \left( \frac{m_{U_3}}{\mu},\frac{M_1}{\mu} \right) \right] 
\nonumber \\
&&- g_t^2 \left[ \,
\frac{3}{4\,\sin^2\beta}\,\ln \frac{\mu^2}{\tm^2}
\,+\,\frac{3}{8}\,\cot^2\beta\, \left( 2\, \ln \frac{\mA^2}{\tm^2} -1\right)  
\right.\nonumber \\
&&~~~~~~~\left.
-\,\frac{\wt{X}_t}{4}\,
\wt{F}_5 \left( \frac{m_{Q_3}}{m_{U_3}} \right)
\,+\,\frac{1}{\sin^2\beta} \wt{F}_6 \left( \frac{m_{Q_3}}{\mu}\right) 
\,+\, \frac{1}{2\,\sin^2\beta}\, \wt{F}_6 \left(
  \frac{m_{U_3}}{\mu} \right)\, \right],\\
\nonumber\\
\label{dgtchi}
(4\pi)^2\,\Delta g_t^{\chi}~~ &=& 
-\frac16\,\tilde{g}_{\text{1u}}\,\tilde{g}_{\text{1d}}
~f \!\left(\frac{M_1}{\mu} \right)
-\frac1{12}\,\left(\tilde{g}_{\text{1u}}^2+\tilde{g}_{\text{1d}}^2\right)
\,\left[g\! \left( \frac{M_1}{\mu} \right)
+3\, \ln \frac{\mu^2}{\tm^2} \right] 
\nonumber\\
&&-\frac12\,\tilde{g}_{\text{2u}}\,\tilde{g}_{\text{2d}}
~f \!\left(\frac{M_2}{\mu} \right)
\,-\,\frac14\,\,\left(\tilde{g}_{\text{2u}}^2+\tilde{g}_{\text{2d}}^2\right)
\,\left[g \!\left( \frac{M_2}{\mu} \right)
+3\, \ln \frac{\mu^2}{\tm^2} \right]~,
\eea
where all loop functions are defined in appendix~\ref{F},
eqs.~(\ref{sys:Fstop})--(\ref{sys:f}).

Once again, in \HSS\ the Higgs-higgsino-gaugino couplings entering
eq.~(\ref{dgtchi}) must be expressed in terms of the gauge couplings
and of $\beta$ by means of eq.~(\ref{matchgt}).  In \sps, on the other
hand, the term $\Delta g_t^{\chi}$ must be removed from the boundary
condition at the scale $\tm$, but it enters the relation between the
top Yukawa coupling of \sps\ and the corresponding coupling of the SM
at the intermediate matching scale:\footnote{Our $\Delta g_t^{\chi}$
  corresponds to the $\tilde \delta_t$ given in eq.~(24) of
  ref.~\cite{Giudice:2011cg}.}
\beq
g_t^{\SM}(\msplit) ~=~ g_t^\split(\msplit)\,\left(1 - \Delta g_t^{\chi}
\right)~.
\eeq

\subsection{Two-loop SUSY-QCD correction to the quartic Higgs
  coupling}
\label{sec:lambda2loop}

To further improve the accuracy of our prediction for the Higgs mass,
we compute the ${\cal O}(g_3^2\,g_t^4)$ two-loop contribution to the
matching condition for the quartic coupling of the light Higgs. Since
there are no WFR contributions to the matching condition at this order
in the couplings, the calculation can be performed entirely in the
effective-potential approach, exploiting the techniques employed in
refs.~\cite{Degrassi:2001yf,Degrassi:2009yq} for the calculation of
the Higgs masses in the MSSM and in the NMSSM.

The ${\cal O}(g_3^2\,g_t^4)$ threshold correction to the light-Higgs
quartic coupling $\lambda$ at the matching scale $\tm$ can be
expressed as
\beq
\label{effpot}
\Delta\lambda^{2\ell} ~ = ~ \frac12\,\left.
\frac{\partial^4 \Delta V^{2\ell,\,\tilde t}}
{\partial^2H^\dagger\partial^2 H}\,\right|_{H=0} 
\!\!+~ \Delta\lambda^{ 2\ell,\,{\rm shift}}~,
\eeq
where $\Delta V^{2\ell,\,\tilde t}$ denotes the contribution
to the MSSM scalar potential from two-loop diagrams involving the
strong gauge interactions of the stop squarks,
\bea
\label{v2as}
\Delta V^{2\ell,\,\tilde t} & = & \frac{g_3^2}{64\,\pi^4}\,
\biggr\{ 2\,\tu\,I(\tu,\tu,0) + 2\,L(\tu,M_3^2,\t)
- 4\,m_t\,M_3\,\sdt\,I(\tu,M_3^2,\t) \nonumber\\
&& +\,\left(1-\frac{\sdt^2}2\right)\,J(\tu,\tu) 
+ \frac{\sdt^2}{2} J(\tu,\td)\;
+ \; \left[ \tul \leftrightarrow \tdl\,,\,
\sdt \rightarrow - \sdt\right] \biggr\}\,,~~
\eea
while $\Delta\lambda^{2\ell,\,{\rm shift}}$ contains additional
two-loop contributions that will be described below.
The loop integrals $I(x,y,z)$, $L(x,y,z)$ and $J(x,y)$ in
eq.~(\ref{v2as}) are defined, e.g., in appendix D of
ref.~\cite{Degrassi:2009yq}, $M_3$ stands for the gluino mass, $\tul$
and $\tdl$ are the two stop-mass eigenstates, and $\sdt \equiv \sin
2\theta_{\tilde t}$, where $ \theta_{\tilde t}$ denotes the stop
mixing angle. The latter is related to the other parameters by
\beq
\label{s2t}
\sin 2\theta_{\tilde t} = \frac{2\, \mt\, (A_t
  -\mu\cot\beta)}{\tu-\td}~.
\eeq

Since we consider scenarios in which electroweak symmetry breaking
(EWSB) occurs only along the direction of the light Higgs doublet $H$,
the calculation of two-loop corrections to its couplings in the
effective-potential approach is considerably simplified with respect
to the MSSM and NMSSM cases. We can express $\tul$, $\tdl$ and $\sdt$
as functions of a field-dependent top mass $m_t = \hat g_t\,|H|$,
where $\hat g_t = \hat y_t\sin\beta$, and re-write eq.~(\ref{effpot})
as
\beq
\label{effpot2}
\Delta\lambda^{2\ell} ~ = ~ \frac{\hat g_t^4}2\,\left(
2\,{\cal D}_2 + 4 \,\mt^2\,{\cal D}_3 + \mt^4\,{\cal D}_4\right) 
\,\Delta V^{2\ell,\,\tilde t}~\biggr|_{\,\mt\rightarrow 0}
\!+~ \Delta\lambda^{2\ell,\,{\rm shift}}~,
\eeq
where we define the operators
\beq
{\cal D}_i ~\equiv~ \left(\frac d{d\mt^2}\right)^i~.
\eeq
We then exploit the following relations for the derivatives of the
field-dependent parameters:
\beq
\label{derivs}
\frac{d m^2_{\tilde t_{1,2}}}{d\mt^2}~=~ 
1 \pm \frac{\sdt}{2\,\mt}\, (A_t -\mu\cot\beta)~,~~~~~~~~
\frac{d \sdt}{d\mt^2}~=~ 
\frac{\sdt}{2\,\mt^2}\,(1-\sdt^2)~.
\eeq

In order to obtain the limit $\mt\rightarrow 0$ in eq.~(\ref{effpot2})
--- of course, {\em after} taking derivatives with respect to
$\mt^2$ --- we use eq.~(\ref{s2t}) to make the dependence of $\sdt$ on
$\mt$ explicit, we expand the function $\Phi(m^2_{\tilde
  t_{i}},M_3^2,\mt^2)$ entering the loop integrals (see appendix D of
ref.~\cite{Degrassi:2009yq}) in powers of $\mt^2$, and finally we
identify $\tul$ and $\tdl$ with the soft SUSY-breaking stop masses
$m_{Q_3}$ and $m_{U_3}$. It turns out that the combination of
derivatives of $\Delta V^{2\ell,\,\tilde t}$ in the right-hand side of
eq.~(\ref{effpot2}) contains terms proportional to $\ln(\mt^2/\tm^2)$,
which diverge for $\mt\rightarrow 0$. However, we must take into
account that above the matching scale $\tm$ the one-loop contribution
to $\lambda$ from the box diagram with a top quark,
\beq
\label{deltatop}
\delta \lambda^{ g_t^4,\, t} = -\frac{3\,\hat g_t^4}{16\pi^2}\,
\left(2\,\ln\frac{\mt^2}{\tm^2} + 3 \right)~,
\eeq
is expressed in terms of the top Yukawa coupling of the MSSM, $\hat
g_t$, whereas below $\tm$ the same contribution is expressed in terms
of the corresponding coupling of the low-energy theory, $g_t$. Being
present both above and below the matching scale, $\delta
\lambda^{g_t^4,\, t}$ does not affect the one-loop threshold
correction to $\lambda$. However, to compute the matching condition at
the two-loop level we must re-express the MSSM coupling entering
$\delta \lambda^{g_t^4,\, t}$ above $\tm$ (including the coupling
implicit in $\mt^2$) according to $\hat g_t \rightarrow g_t \,(1+
\Delta g_t^{\phi,\,g_s^2})\,$, where $\Delta g_t^{\phi,\,g_s^2}$
denotes the terms proportional to $g_s^2$ in eq.~(\ref{dgtphi}).  This
induces a two-loop contribution to $\Delta\lambda^{2\ell}$ which
cancels out the terms proportional to $\ln(\mt^2/\tm^2)$ in the
derivatives of the effective potential. In addition, we re-express the
MSSM coupling entering the terms proportional to $\hat g_t^4$ in the
one-loop stop contribution to $\lambda$, see eq.~(\ref{threshsusy}),
according to $\hat g_t \rightarrow g_t \,(1+ \Delta g_t^{\phi,\,g_s^2}
+ \Delta g_t^{{\rm ren},\,g_s^2}\,)\,$. The correction $\Delta
g_t^{{\rm ren},\,g_s^2}$ denotes the term proportional to $g_s^2$ in
eq.~(\ref{dgtreg}), and accounts for the fact that we renormalize the
couplings of the low-energy theory in the $\msbar$ scheme, while the
effective-potential calculation of the two-loop contributions to
$\lambda$ was performed in the $\drbar$ scheme. The combined effect of
these shifts is the term denoted as $\Delta\lambda^{2\ell,\,{\rm
    shift}}$ in eqs.~(\ref{effpot}) and (\ref{effpot2}). Note that the
$\drbar\,$--$\,\msbar$ redefinition of the Yukawa coupling in the one-loop
top contribution $\delta \lambda^{g_t^4,\, t}$ has the same effect
above and below the matching scale, therefore it does not contribute
to $\Delta\lambda^{2\ell,\,{\rm shift}}$.
 
It is interesting to remark that the two-loop contributions arising
from the operators ${\cal D}_3$ and ${\cal D}_4$ in
eq.~(\ref{effpot2}) cancel out completely against the shift induced
when the corresponding contributions in the one-loop part --- in
practice, the non-logarithmic term in $\delta \lambda^{g_t^4,\, t}$,
see eq.~(\ref{deltatop}) --- are expressed in terms of the top Yukawa
coupling of the low-energy theory. Consequently, the final result for
$\Delta\lambda^{2\ell}$ originates only from the operator ${\cal
  D}_2$, and is therefore proportional to the stop contribution to the
${\cal O}(g_3^2\,g_t^2\,\mt^2)$ correction to the light-Higgs mass in
the MSSM. This ``decoupling'' property of the two-loop SUSY
contribution to the light-Higgs quartic coupling was also noted, in a
slightly different context, in ref.~\cite{Brucherseifer:2013qva}.

In the \spsh\ case, we can take the limit of vanishing gluino mass in
the two-loop correction to the Higgs quartic coupling. We
obtain\,\footnote{We henceforth drop the distinction between $\hat
  g_t$ and $g_t$ in $\Delta\lambda^{2\ell}$, because it amounts to a
  higher-order effect.}
\bea
\Delta\lambda^{2\ell}&=& -\frac{g_3^2\,g_t^4}{32\,\pi^4}~\biggr\{
3  + 4\, \ln\xqu + 8\, \ln^2 \xqu + 6\,\ln^2\frac{\Q}{\tm^2}
-4 \,(1+ 3\ln\xqu)\,\ln\frac{\Q}{\tm^2} \nonumber\\
&&~~~~~~~~~
+ \xtt\,\left[
\frac{12\,\xqu\,\ln\xqu}{\xqu^2-1}\,\left(2\,\ln\frac{\Q}{\tm^2}-1\right)
-\frac{16\,\xqu\,(\xqu^2-2)\,\ln^2\xqu}{(\xqu^2-1)^2}\,\right]\nonumber\\
&&~~~~~~~~~
+ \xtt^2\,\left[
\frac{6\,\xqu^2\,(5+\xqu^2)\,\ln\xqu}{(\xqu^2-1)^3}
+\frac{4\,\xqu^2\,(\xqu^4-4\,\xqu^2-5)\,\ln^2\xqu}{(\xqu^2-1)^4}
\right.\nonumber\\
&&~~~~~~~~~~~~~~~~~
\left.-\frac{10\,\xqu^2}{(\xqu^2-1)^2}
+ \frac{12\,\xqu^2}{(\xqu^2-1)^2}
\left(1-\frac{\xqu^2+1}{\xqu^2-1}\,\ln\xqu\right)\,\ln\frac{\Q}{\tm^2}\,
\right]\biggr\}~,
\eea
which for equal stop masses $m_{Q_3}=m_{U_3}=\MS$ reduces to
\beq
\Delta\lambda^{2\ell}~=~
-\frac{g_3^2\,g_t^4}{32\,\pi^4}~\left[
3-2\,\xtt+\frac{\xtt^2}6 
%- (4-12\,\xtt+\xtt^2)\,\ln\frac{\MS^2}{\tm^2}
%+ 6\,\ln^2\frac{\MS^2}{\tm^2}
\right]~.
\eeq

In the case of \HSS, on the other hand, we cannot consider the gluino
mass much smaller than the stop masses.  The formula for
$\Delta\lambda^{2\ell, {\rm {\scriptscriptstyle HSS}}}$ with full
dependence on $M_3$, $m_{Q_3}$ and $m_{U_3}$ is lengthy and not
particularly illuminating, but in the limit $M_3 =m_{Q_3} = m_{U_3} =
\MS$ it simplifies to
\bea
\label{d2lhss}
\Delta\lambda^{2\ell, {\rm {\scriptscriptstyle HSS}}}
&=& \frac{g_3^2\,g_t^4}{96\,\pi^4}~\left[
-12\,\frac{X_t}{\MS}-6\,\frac{X_t^2}{\MS^2}
+14\,\frac{X_t^3}{\MS^3}+\frac12\,\frac{X_t^4}{\MS^4}-\frac{X_t^5}{\MS^5}
\right]\,,
%&&~~~~~~~~~~\left.
%+\left(24\,\frac{X_t}{\MS}-24\,\frac{X_t^2}{\MS^2}-4\,\frac{X_t^3}{\MS^3}
%+\frac{X_t^4}{\MS^4}\right)
%\,\ln\frac{\MS^2}{\tm^2}-18\,\ln^2\frac{\MS^2}{\tm^2}\,\right]\,.
\eea
It is easy to check that, consistently with the ``decoupling''
behavior discussed above, the ${\cal O}(g_3^2\,g_t^4)$ threshold
correction to the light-Higgs quartic coupling in eq.~(\ref{d2lhss})
could be recovered directly from the known results for the ${\cal
  O}(g_3^2\,g_t^2\,\mt^2)$ correction to the light-Higgs mass. In
particular, it is sufficient to subtract the top-quark contribution
given, e.g., in eq.~(20) of ref.~\cite{DDEEGIS} from the full MSSM
correction given, e.g., in eq.~(21) of ref.~\cite{Espinosa:1999zm} (in
the latter it is also necessary to transform the $\drbar$ top mass of
the MSSM, denoted by $\mt$, into the $\msbar$ top mass of the SM,
denoted by $\overline m_t$).

\section{The Higgs mass and supersymmetry}\label{sec:hiSUSY}

\subsection{Quasi-natural SUSY}\label{sec:QN}
In ``quasi-natural'' supersymmetry all superparticles have masses
$\tm$ in the range between a few to tens of TeV.  A combination of
SUSY-breaking parameters must be fine-tuned at 1 part in $(\tm
/M_Z)^2$ in order to achieve the correct $Z$-boson mass.  In such
scenarios, a fixed-order calculation of the MSSM prediction for the
Higgs mass is no longer accurate, because corrections enhanced by $\ln
(\tm/M_Z)$ must be resummed.  This can be done with the strategy of
ref.~\cite{Giudice:2011cg}, including now higher-order corrections:

\begin{enumerate}
\item We assume that physics at the weak scale is described by the SM,
  and extract from data the $\msbar$-renormalized parameters with
  two-loop precision in all couplings, adopting the results
  of~\cite{SMpar}.\footnote{Three-loop QCD corrections to $g_t$ and to
    $\lambda$ are also partially available and confirm the estimated
    higher-order uncertainties.}

\item We evolve the SM parameters from the weak scale up to the SUSY
  scale $\tm$ using the known RGEs of
  the SM at three loops.

\item At $\tm$ we equate the quartic Higgs coupling $\lambda$ with
  its supersymmetric prediction, as computed in section~\ref{sec:thre}
  including all superparticle thresholds at one loop, and the QCD
  superparticle thresholds at two loops.
\end{enumerate}

Depending on the specific analysis being performed, the third step
either determines one of the SUSY parameters (e.g., $\tan\beta$) at
the scale $\tm$ or determines the physical Higgs mass corresponding to
a given set of SUSY parameters (in this case, the input Higgs mass in
the first step is varied until the value of $\lambda(\tm)$ obtained by
RG evolution matches the SUSY prediction).

\begin{figure}[t]
$$\includegraphics[width=0.75\textwidth]{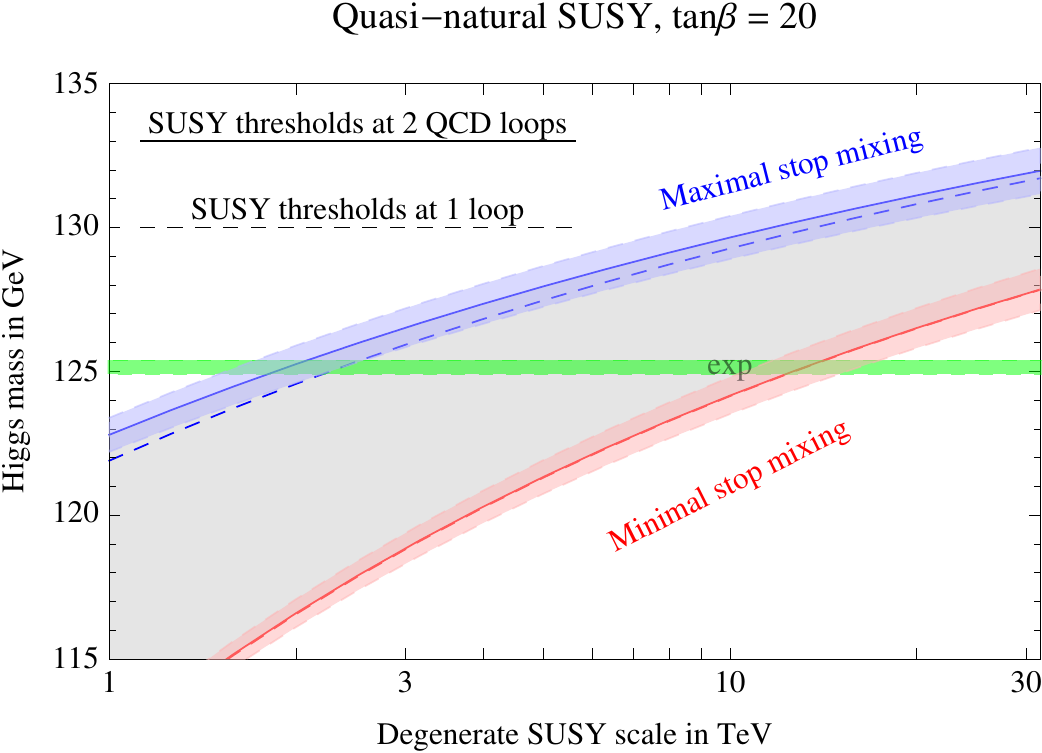}$$
\caption{\em The Higgs mass as a function of the SUSY scale,
  with a degenerate spectrum of superparticles and $\tan\beta=20$.  
  We vary the stop mixing
  parameter $X_t$ in such a way as to obtain minimal $\MH$ (red lines) and
  maximal $\MH$ (blue lines). The
  solid (dashed) lines include (neglect) the effect of the two-loop
  SUSY-QCD corrections to $\lambda$. The solid and dashed red lines overlap. 
  The red and blue bands around the solid
  lines indicate the uncertainty associated to the measurement of the SM 
  input parameters. The green band indicates the measured Higgs mass.
  \label{fig:Mhpred}}
\end{figure}

In figure\fig{Mhpred} we consider a simplified scenario with
$\tan\beta=20$ and a degenerate superparticle spectrum ({\it i.e.},
all SUSY mass parameters, including $\mA$ and $\mu$, equal to a common
mass $\MS$), and show the Higgs mass as a function of $\MS$. For a
given value of $\MS$, we vary the stop mixing parameter $X_t =
A_t-\mu\cot\beta$ to obtain the minimal (red lines) and maximal (blue
lines) values of the Higgs mass. The former are obtained in the
vicinity of $X_t =0$, and the latter in the vicinity of $X_t =
\sqrt{6}\,\MS$, {\it i.e.}, the value that maximizes the dominant
${\cal O}(g_t^4)$ threshold correction to $\lambda(\tm)$ in
eq.~(\ref{threshsusy}). In both the minimal- and maximal-mixing cases,
the solid line includes the effect of the two-loop SUSY-QCD
corrections to $\lambda(\tm)$ described in section
\ref{sec:lambda2loop}, while the dashed line does not include it. The
bands around the solid lines represent the parametric uncertainty of
the prediction for the Higgs mass, obtained by varying the pole top
mass and the strong gauge coupling within their $1\sigma$ experimental
uncertainty, $M_t = 173.34\pm 0.76\GeV$~\cite{topmass} and
$\alpha_3(M_Z) = 0.1184\pm 0.0007$~\cite{alpha3}. The green horizontal
band indicates the measured value for the Higgs mass, $\MH^{\rm exp}
=\Mhexp$, obtained from a naive average of the ATLAS and CMS
results~\cite{HiggsMass}.

The figure suggests that, for moderately large $\tan\beta$ and
degenerate SUSY masses, a value of $\tm$ around 2~TeV is needed to
predict a Higgs mass compatible with the experimental result, as long
as the Higgs-stop coupling $X_t$ is adjusted to maximize the
correction. In the case of vanishing $X_t$, on the other hand, SUSY
masses greater than 10 TeV are needed. In both cases, a wider range of
values of $\MS$ becomes acceptable when the experimental uncertainty
of $M_t$ and $\alpha_3$ is taken into account. Lowering $\tan\beta$
would reduce the tree-level part of the boundary condition for
$\lambda$, requiring even larger SUSY masses.

The comparison between the solid and dashed blue lines in
figure\fig{Mhpred} shows that, for large $X_t$, the two-loop ${\cal
  O}(g_3^2\,g_t^4)$ corrections to $\lambda(\tm)$ can increase the
Higgs mass by up to 1~GeV at low $\MS$, but their effect is reduced as
$\MS$ gets larger (indeed, both $g_t$ and $g_3$ decrease at higher
scales). On the other hand, eq.~(\ref{d2lhss}) shows that those
corrections vanish for $X_t=0$ and degenerate SUSY masses. 
% when the matching scale is also chosen as $\tm = \MS$. 
Consequently, the solid and dashed red lines overlap in the figure.

\subsubsection*{Comparison with other recent computations}

It is useful to compare our results for the Higgs mass with those in
two recent papers~\cite{Wagner,Hollik} where the importance of
resumming the large logarithms in heavy-SUSY scenarios was
emphasized.\footnote{An earlier study of heavy-SUSY scenarios,
  ref.~\cite{Feng}, neglected the resummation of large logarithms,
  thus overestimating the Higgs mass by more than 10~GeV for stop
  masses around 10~TeV.}  The renormalization-group (RG) calculation
in ref.~\cite{Wagner} is conceptually similar to ours, although the SM
relation between the running quartic coupling and the pole Higgs mass
in step~1 is computed only at one loop, and two-loop terms of ${\cal
  O}(g_t^6)$, which we neglect, are included in the SUSY correction to
$\lambda(\tm)$ in step~3. Ref.~\cite{Hollik}, on the other hand,
combines the ``diagrammatic'' calculation of the MSSM Higgs masses
implemented in the code {\tt FeynHiggs}~\cite{feynhiggs} --- which
includes full one-loop~\cite{FH-oneloop} plus dominant
two-loop~\cite{Degrassi:2001yf,FH-twoloop} corrections --- with a
resummation of the leading and next-to-leading logarithmic terms
controlled exclusively by $g_t$ and $g_3$.

We again focus on a simplified scenario with heavy and degenerate SUSY
masses, $\MS=10~\TeV$, and take $X_t=0$ and $\tan\beta=20$. Fixing the
SM input parameters to their central values, we find $\MH =
123.6$~GeV, which should be compared to the value $\MH = 123.2$~GeV in
the upper-left plot of figure~1 in ref.~\cite{Wagner}, and to the
value $\MH = 126.5$~GeV obtained with the version of {\tt FeynHiggs}
described in ref.~\cite{Hollik}.\footnote{To perform the comparison,
  we converted the $\drbar$ input parameters $\MS$ and $X_t$ to the
  ``on-shell'' scheme adopted by {\tt FeynHiggs}, using results from
  ref.~\cite{Espinosa:2000df}. However, in this point the effect of
  the conversion on the Higgs mass amounts only to a few hundred MeV.}
While the agreement between our result and the one of
ref.~\cite{Wagner} appears satisfactory in view of the small
differences between the two RG calculations, the $\sim\!3$~GeV
discrepancy with the ``hybrid'' (i.e., diagrammatic+RG) calculation of
ref.~\cite{Hollik} deserves further discussion.

A decade ago, the theoretical uncertainty of partial two-loop
calculations of the MSSM Higgs mass such as the one implemented in
{\tt FeynHiggs} was indeed estimated to be of the order of
3~GeV~\cite{Degrassi:2002fi,Allanach:2004rh}. However, that estimate
was developed for fixed-order calculations in what were then
considered natural regions of the MSSM parameter space, and it does
not necessarily apply to RG calculations in heavy-SUSY scenarios.
A realistic assessment of the theoretical uncertainty of our
Higgs-mass calculation should take into account three sources of
uncertainty: the first are missing higher-order terms in the SM
computations of steps 1 and 2, which were estimated in
ref.~\cite{SMpar} to induce an uncertainty of $\pm 0.2$~GeV in the
Higgs mass. The second are missing higher-order corrections in the
SUSY thresholds of step 3: by varying the matching scale by a factor
of 2 around $\MS$, we estimate that these missing corrections induce
an uncertainty of $\pm 0.5$~GeV in the Higgs mass. Indeed, we would
not expect their effect to be much larger than the one of the known
two-loop ${\cal O}(g_3^2\,g_t^4)$ corrections, which, even for large
stop mixing, shift the Higgs mass by at most 0.4~GeV in the scenario
with SUSY masses all equal to $\MS=10$~TeV. Finally, a third source of
uncertainty are effects suppressed by powers of $v^2/\MS^2$ and by a
loop factor, which arise because in steps 1 and 2 we employ the SM as
an effective theory, thus neglecting heavy-superparticle effects in
the determination of the running couplings, and because in step 3 we
neglect the effects of EWSB when matching the MSSM couplings onto the
SM ones. Of course, the relevance of ${\cal O}(v^2/\MS^2)$ effects
decreases for increasing superparticle masses: we estimate that for
$\MS=10$~TeV the uncertainty in the Higgs mass induced by those
effects is already negligible.
Putting all together, the theoretical uncertainty of our result for
the Higgs mass in the point with $\MS=10$~TeV, $X_t=0$ and
$\tan\beta=20$ should not be larger than $\pm 1\GeV$, which makes it
incompatible with the corresponding result of ref.~\cite{Hollik}. The
observed 3-GeV discrepancy might be explained by the fact that the
resummation procedure in ref.~\cite{Hollik} covers only a subset of
the leading and next-to-leading $\ln(\MS/M_t)$ effects, thus the
calculation of the Higgs mass is still affected by residual large
logarithms (e.g., those controlled by the electroweak gauge
couplings).

Of course, the impact of the second and third sources of uncertainty
discussed above depends strongly on the considered point in the MSSM
parameter space. Higher-order effects in the threshold corrections at
the matching scale might become more relevant for non-degenerate SUSY
masses or for lower $\tm$ (where the couplings $g_t$ and $g_3$ are
larger). Also, the effects suppressed by the superparticle masses
become obviously larger for lower $\MS$. In particular, for SUSY mass
parameters all equal to $\MS=1$~TeV we can expect the RG resummation
to play a lesser role in the accuracy of the Higgs-mass calculation,
while the corrections that we neglect become more relevant. Taking
again $\tan\beta=20$ and varying $X_t$, we find a maximal Higgs mass
$\MH^{\rm max} \approx 123$~GeV in this scenario (see the ``maximal
mixing'' line on the left edge of figure~\ref{fig:Mhpred}). In
contrast, {\tt FeynHiggs} predicts $\MH^{\rm max} \approx 129 -
131$~GeV (depending on the code's settings), while codes such as {\tt
  SoftSusy}~\cite{Allanach:2001kg}, {\tt
  SuSpect}~\cite{Djouadi:2002ze} and {\tt SPheno}~\cite{Porod:2003um},
which compute the mass spectrum of the MSSM including the full
one-loop and dominant two-loop corrections to the Higgs masses in the
$\drbar$ scheme,\footnote{\,While these three codes implement the same
  corrections to the Higgs masses, they differ in the determination of
  the running couplings.}  predict $\MH^{\rm max} \approx 124.5 -
126.5$~GeV. Such a spread in the Higgs-mass predictions --- in a
scenario where there is no obvious argument to favor one calculational
approach over the others --- points to a large theoretical
uncertainty, and to the need of improving the calculation with the
inclusion of higher-order effects.

\subsection{\HSS}
\label{sec:high}

In \HSS, all supersymmetric masses lie around the same scale $\tm$,
which can be much larger than the weak scale.  The measured Higgs mass
$\MH^{\rm exp} =\Mhexp$ is reproduced in a band of the
($\tm,\,\tan\beta)$ plane, as discussed in ref.~\cite{Giudice:2011cg}
(see also refs.~\cite{Hall:2009nd, Cabrera:2011bi, Arbey:2011ab,
  Ibanez:2013gf, Hebecker:2013lha, DDEEGIS,Delgado:2013gza,Wagner}).
Here we update the analysis of ref.~\cite{Giudice:2011cg}, including
our improved calculation of the supersymmetric threshold corrections
discussed in section~\ref{sec:thre}.

\begin{figure}[t]
$$\includegraphics[width=0.45\textwidth]{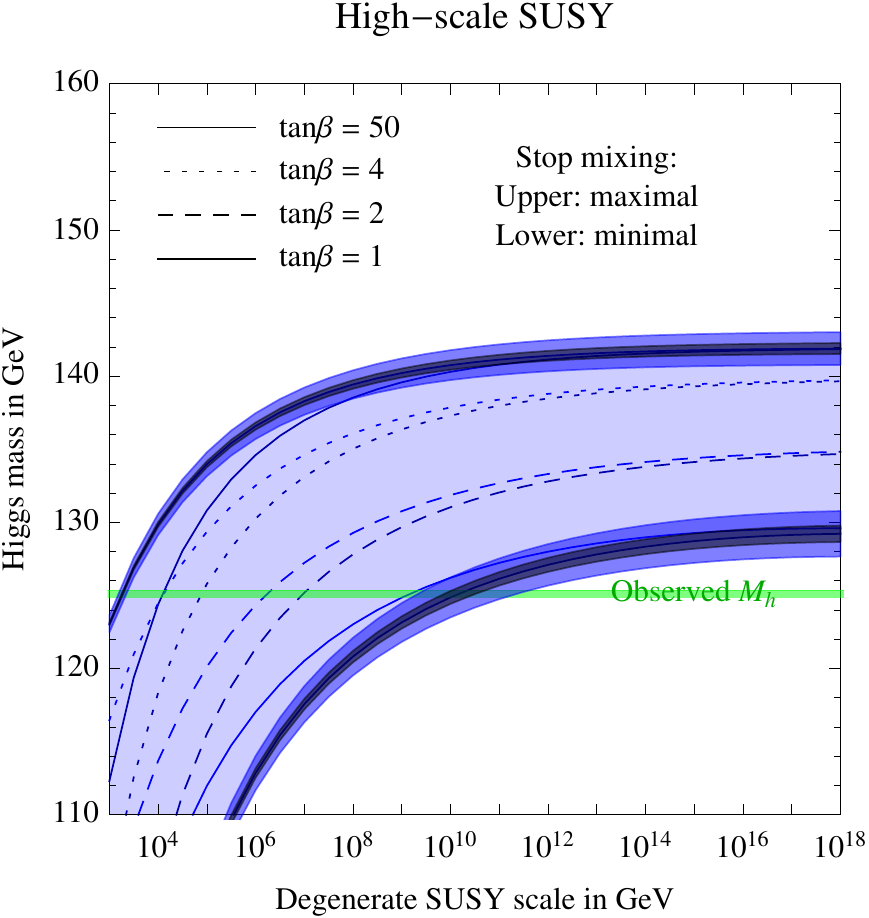}\qquad
\includegraphics[width=0.45\textwidth]{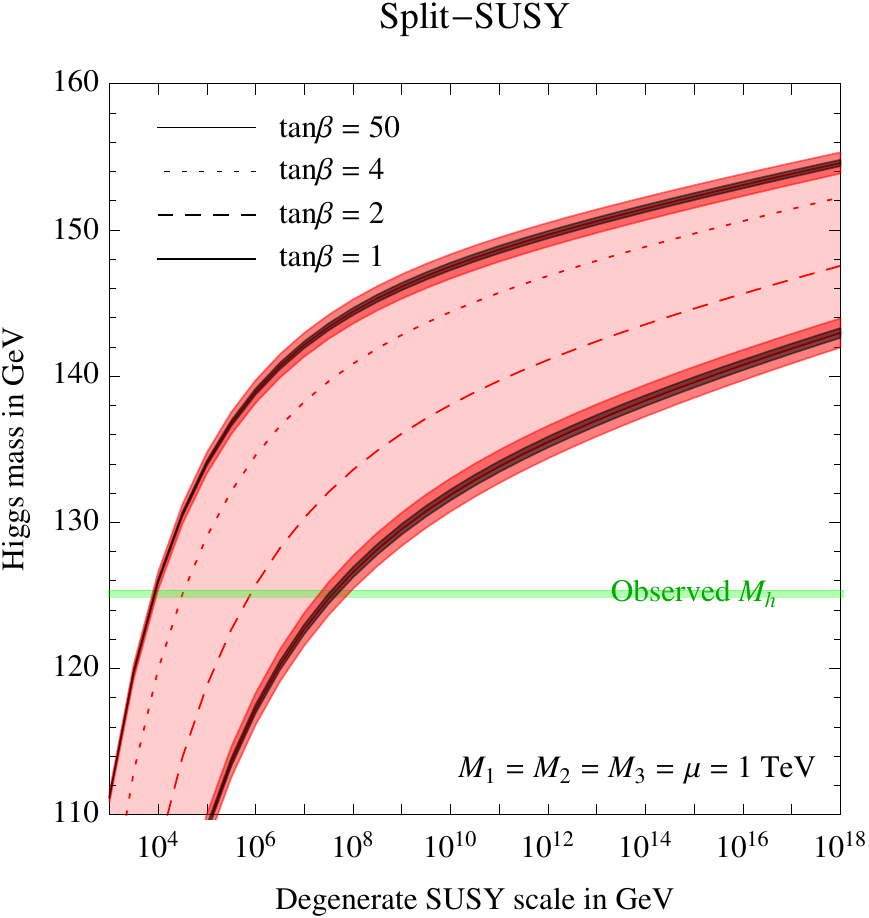}$$
\caption{\em {\bf Left:} the Higgs mass as a function of the SUSY
  scale $\tm$, with a degenerate spectrum of superparticles. We vary
  the Higgs-stop coupling $A_t$ in such a way as to obtain minimal
  $\MH$ (lower lines) and maximal $\MH$ (upper lines) at fixed
  $\tan\beta=\{1,2,4,50\}$. The bands around the extremal solid lines
  are obtained from $1\sigma$ variations of $\alpha_3(M_Z)$ (thinner
  band in gray) and $M_t$ (larger band in color).  The green
  horizontal band indicates the measured Higgs mass. {\bf Right:} same
  as in the left plot, for a split spectrum with gaugino and higgsino
  masses set to $1\TeV$ and with
  $A_t=0$.}
\label{fig:HeavySUSY}
\end{figure}

The left plot in figure\fig{HeavySUSY} shows our updated result.  We
assume a degenerate spectrum with all superparticle masses set equal
to $\tm$, and plot our prediction for the Higgs mass $\MH$ as function
of $\tm$ for $\tan\beta=\{1,2,4,50\}$, varying the soft SUSY-breaking
Higgs-stop coupling $A_t$ in order to minimize or maximize $\MH$.  We
also plot the uncertainty on the prediction for $\MH$ induced by the
experimental uncertainty on the SM input parameters $M_t$ and
$\alpha_3(M_Z)$. The plot shows that, even allowing for a $1\sigma$
reduction in the pole top mass, the measured value of the Higgs mass
cannot be reproduced in this simplified scenario if the common SUSY
scale $\tm$ is larger than roughly $3\times 10^{11}\GeV$.  However,
this upper bound on $\tm$ is very sensitive to the top mass and
completely evaporates if $M_t$ is reduced within its $3\sigma$ range.

\begin{figure}[t]
$$
\includegraphics[width=0.65\textwidth]{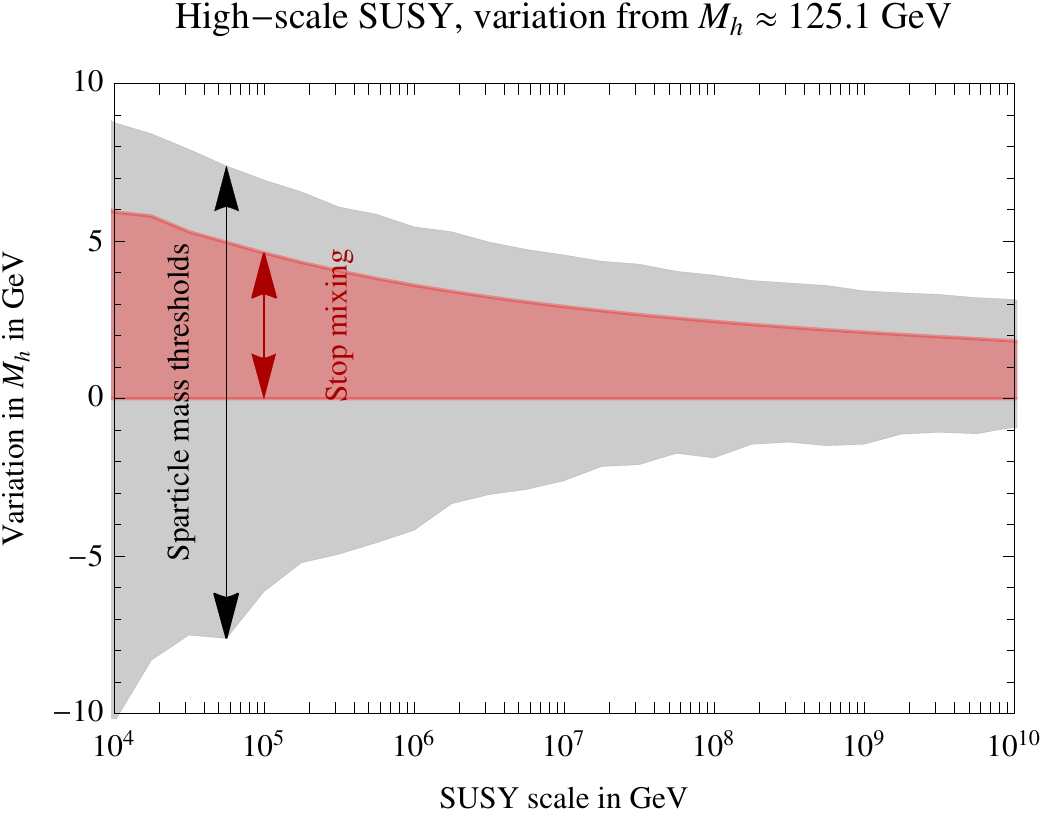}
$$
\caption{\em Variation in the prediction for $\MH$ in \HSS\ from
  random scanning each superparticle mass parameter $M_1, M_2, M_3,$
  $m_{Q_i}, m_{U_i}, m_{D_i}, m_{E_i}, m_{L_i}$ and $\mu$ up to a
  factor 3 above or below the scale $\tm$, and the Higgs-stop coupling
  $A_t$ within the range allowed by vacuum stability. The darker (red)
  band shows the variation due only to $A_t$.
\label{fig:Mhband}}
\end{figure}

Next, we consider non-degenerate superparticle spectra.  Given that
superparticle masses are unknown, we randomly scan over them, varying
independently the mass parameters $M_1$, $M_2$, $M_3$, $m_{Q_i},
m_{U_i}, m_{D_i}, m_{E_i}, m_{L_i}$ (distinguishing the third
generation from the other two) and $\mu$ between $\tm/3$ and $3\,\tm$,
and the Higgs-stop coupling $A_t$ within the range allowed by vacuum
stability (see next subsection). Figure\fig{Mhband} shows the induced
variation in $\MH$ with respect to the prediction obtained with
degenerate superparticles at a given mass $\tm$, and with $\tan\beta$
and $A_t$ adjusted so that $X_t=0$ and $\MH=\Mhnoerr$ (this restricts
our scan to the range $10^4~{\rm GeV}\, \simlt\, \tm\,\simlt\,
10^{10}~{\rm GeV}$, where the measured value of the Higgs mass can be
reproduced with central values of the SM input parameters). The darker
(red) region in figure\fig{Mhband} denotes the effect of varying only
$A_t$.  The variation in $\MH$ is maximal ($\approx 10\GeV$) in the
case of quasi-natural SUSY, $\tm\approx10^{4}\GeV$, where $g_t$ and
$g_3$ are large and induce sizable threshold corrections. The
variation in $\MH$ rapidly decreases with increasing $\tm$, going down
to about 2--$3\GeV$ at the scale $\tm\approx10^{10}\GeV$. This shows
that the prediction of $\MH$ becomes more robust against unknown
supersymmetric threshold corrections as one considers larger values of
$\tm$.

\begin{figure}[t]
$$
\includegraphics[width=0.45\textwidth]{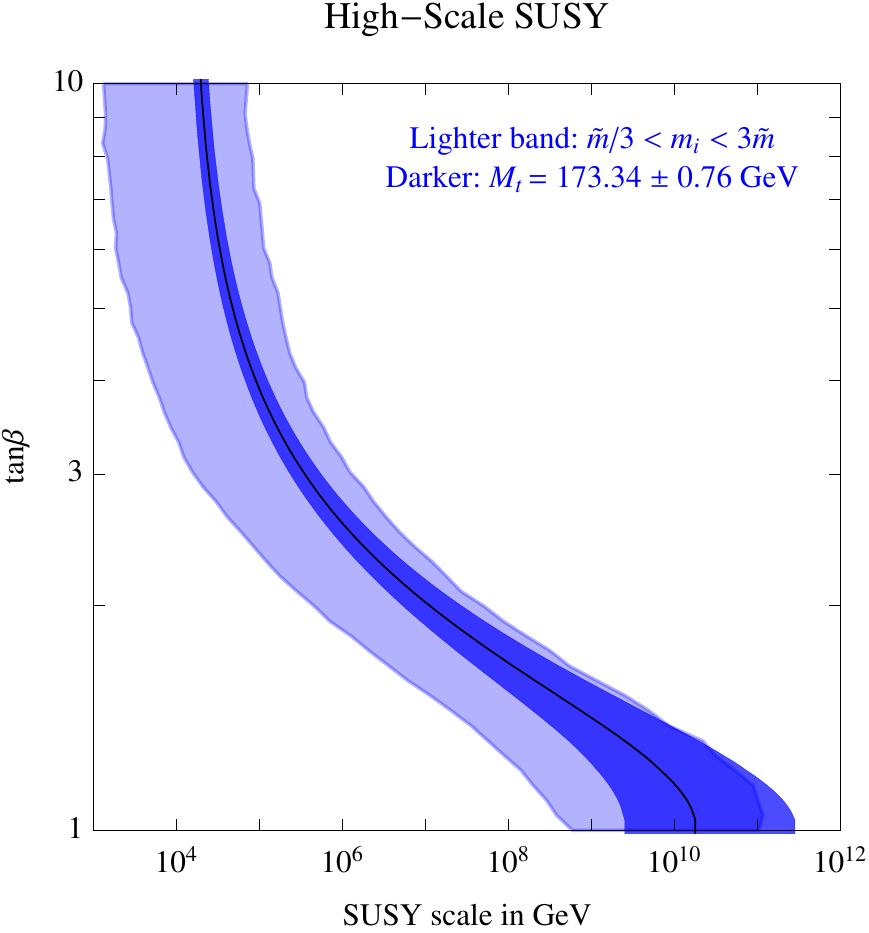}
\qquad\includegraphics[width=0.45\textwidth]{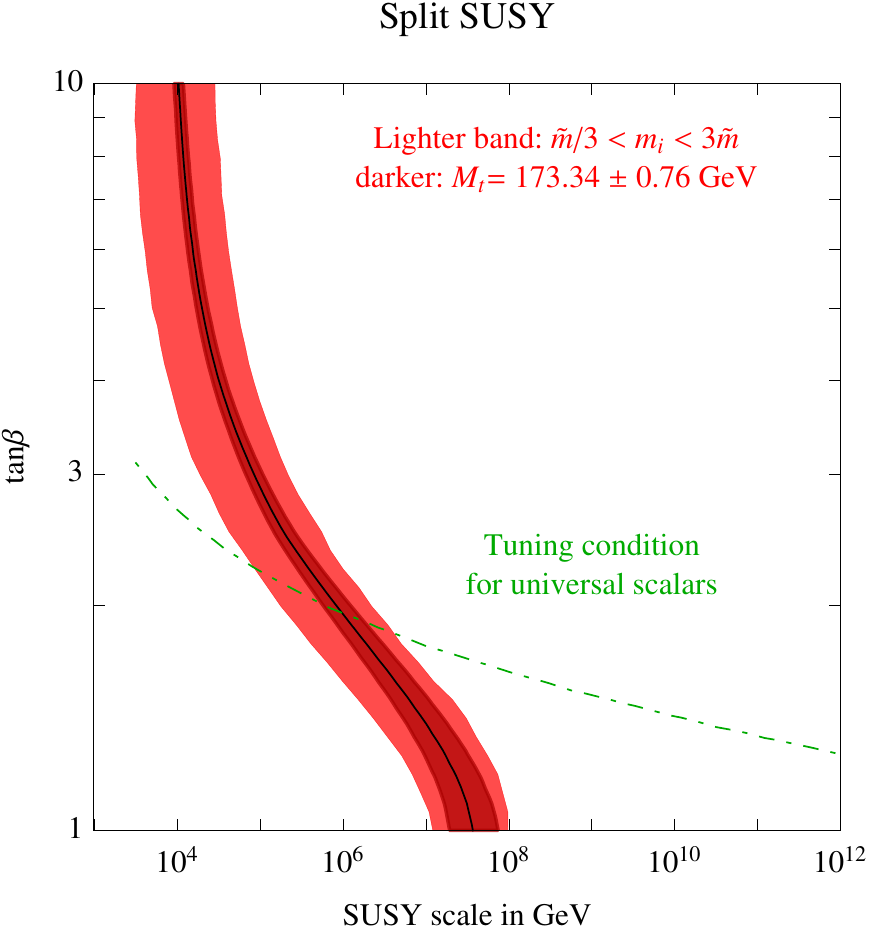}
$$
\caption{\em {\bf Left:} Regions in the $(\tm,\tan\beta)$ plane that
  reproduce the observed Higgs mass for \HSS.  The black solid line
  gives the prediction for $X_t=0$, mass-degenerate superparticles,
  and central values for the SM parameters.  The light-blue band shows
  the effect of superparticle thresholds by varying the supersymmetric
  parameters $M_1, M_2, M_3,$ $m_{Q_i}, m_{U_i}, m_{D_i}, m_{E_i}$,
  $m_{L_i}$ and $\mu$ randomly by up to a factor 3 above or below the
  scale $\tm$, and $A_t$ within the range allowed by vacuum stability.
  The dark-blue band corresponds to mass-degenerate superparticles,
  but includes a 1$\sigma$ variation in $M_t$. {\bf Right:} Same as
  the left plot for the case of \sps. The gaugino and higgsino masses
  are all set to $1\TeV$, and $A_t=0$. The dot-dashed curve corresponds
  to the EW tuning condition in the case of universal scalar masses at
  the GUT scale.
  \label{fig:deltalambdaSUSY}}
\end{figure}

Finally, the left plot in figure\fig{deltalambdaSUSY} shows the region
in the $(\tm, \tan\beta)$ plane where the measured Higgs mass is
reproduced in \HSS.  The solid black curve is the prediction obtained
with $X_t=0$ and exact mass degeneracy at the scale $\tm$ of all
supersymmetric particles, assuming central values of the SM input
parameters $\MH, M_t$ and $\alpha_3(M_Z)$. The effect of a $1\sigma$
variation of the top pole mass is illustrated by the dark blue
band. At low $\tan\beta$, corresponding to large $\tm$, the variation
of $M_t$ strongly affects $\tm$. This is mainly because, as shown in
figure\fig{HeavySUSY}, the dependence of $\MH$ on $\tm$ becomes rather
flat in \HSS\ when $\tm \simgt 10^9$~GeV. Therefore, a small change in
$\MH$ implies a large change in $\tm$.  The light blue band shows the
effect of varying independently all the supersymmetric mass parameters
between $\tm/3$ and $3\,\tm$, and $A_t$ within the range allowed by
vacuum stability, as in figure\fig{Mhband}. Unlike the case of $M_t$,
the impact of supersymmetric thresholds in the extraction of $\tm$
does not show a strong dependence on $\tm$ itself. This is due an
approximate cancellation between two opposing effects: on one hand, as
mentioned above, at large $\tm$ any change in the prediction of $\MH$
(whether from $M_t$ or from supersymmetric thresholds) has an
amplified impact on the determination of $\tm$. On the other hand,
supersymmetric thresholds are smaller at large $\tm$ (see
figure\fig{Mhband}). The two effects nearly compensate each other, and
the impact of supersymmetric thresholds on the light-blue band in
figure\fig{deltalambdaSUSY} is fairly uniform.
 
The left plot in figure\fig{deltalambdaSUSY} shows again how the Higgs
mass measurement implies an upper bound on $\tm$ of about
  $2\times10^{10}\GeV$ in \HSS\ with degenerate supersymmetric
masses and central values of the SM parameters. This bound can be
raised to about $10^{11}\GeV$ if the supersymmetric square
masses differ by about one order of magnitude. So it is difficult for
supersymmetric thresholds to raise the bound up to the Planck (or GUT)
scale, unless, as previously noticed, $M_t$ is  $3\sigma$ below its central value.

As a side remark we note that, in the region of interest, the MSSM top
Yukawa coupling $y_t$ always remains perturbative. The condition that
there are no Landau poles below the Planck scale implies $\tm\simgt
10^7$~GeV for $\tan\beta=1$. This constraint is easily satisfied by
the band in the left plot of figure\fig{deltalambdaSUSY}.

\subsubsection*{Vacuum stability in \HSS}

Our scans of the SUSY parameter space are restricted to spectra that
satisfy the vacuum stability condition.  This is an important issue,
because this condition eliminates spurious corrections that could
reduce the Higgs mass when the parameter $\tilde X_t = (A_t -
\mu\cot\beta)^2/m_{Q_3}m_{U_3}$ is larger than about 12.  The
well-known bounds valid in the case of natural SUSY (see, e.g.,
ref.~\cite{Casas:1995pd}) need to be adapted to the case of \HSS,
where the mass term for a combination of the two MSSM Higgs doublets
almost vanishes because of the electroweak fine-tuning. In order to
determine the upper bound on $\tilde X_t$, let us consider the scalar
potential for the stop-Higgs system
\bea V &=& m_{Q_3}^2 |\tilde Q_3|^2 + m_{U_3}^2 |\tilde U_3|^2 
+ \frac{g_t}{\sin \beta}
\left( A_t H_u \tilde Q_3 \tilde U_3+\mu H_d^*\tilde Q_3 \tilde U_3
+\hbox{h.c.}\right)
\nonumber \\
&+& \frac{g_t^2}{\sin^2\beta} \left(|H_u \tilde Q_3|^2 + |H_u\tilde U_3|^2 +
|\tilde Q_3 \tilde U_3|^2 \right) 
+ \hbox{Higgs-mass terms} + D\hbox{-terms} \ ,
\eea
where the appropriate $\SU(2)_L$ contractions are implicit and where
$g_t$ is the top Yukawa coupling of the SM. Let us consider the
potential along the direction of the approximately-massless Higgs
field $H$ (with $H_u = H \sin\beta$, $H_d = \epsilon H^* \cos\beta$)
and along a squark direction such that the $D$-terms vanish. We
parameterize this $D$-flat direction with a real field $\phi$, defined
by
\beq
H_u^i=\frac{\phi \sin\beta}{\sqrt{2}}\delta^i_2
,~~~~H_d^i=\frac{\phi \cos\beta}{\sqrt{2}}\delta^i_1
,~~~~ \tilde Q_3^{i\alpha}=\phi \ 
\sqrt{\frac{|\cos 2 \beta |}{2}}\ \delta^i_1\delta^\alpha_1
,~~~~ \tilde U_3^\alpha=\phi \ 
\sqrt{\frac{|\cos 2 \beta |}{2}}\ \delta^\alpha_1 \, ,
\eeq
where $i$ and $\alpha$ are weak $\SU(2)_L$ and color $\SU(3)_c$
indices, respectively.  The potential for $\phi$ becomes
\beq
V= |\cos 2 \beta | \left[ \left( m_{Q_3}^2 + m_{U_3}^2 \right) \frac{\phi^2}{2}
-\frac{g_t}{\sqrt{2}}\left( A_t -\frac{\mu}{\tan\beta}\right) \phi^3
+g_t^2 \left( 1-\frac{1}{4\sin^2\beta}\right) \phi^4 \right] \ .
\eeq
The requirement that the color-breaking minimum $\vev{\phi} \ne 0$ is
not deeper than the electroweak minimum finally implies
\beq 
\tilde{X}_t= \frac{(A_t-\mu\cot\beta)^2}{m_{Q_3}m_{U_3}}
< \left( 4-\frac{1}{\sin^2\beta}\right)
\left(\frac{m_{Q_3}}{m_{U_3}}+\frac{m_{U_3}}{m_{Q_3}}\right) \ .
\eeq
This constraint has been used to derive the bands in
figure\fig{Mhband} and in the left plot of
figure\fig{deltalambdaSUSY}.

\subsubsection*{Gauge-coupling unification}

\begin{figure}[t]
$$
\includegraphics[width=0.65\textwidth]{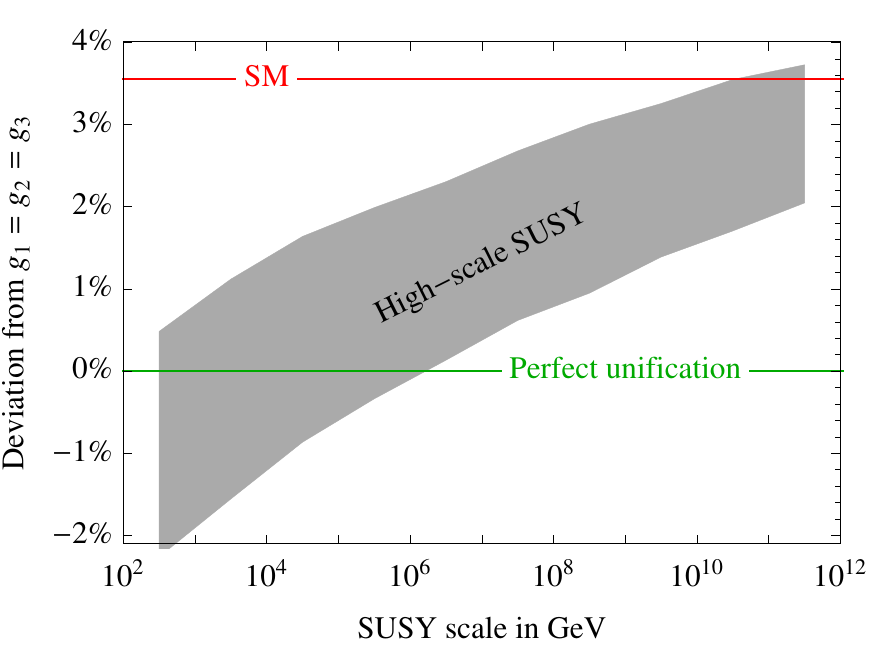}$$
\caption{\em The minimum amount (in percent) by which the unification
  relation $\hat g_{1,2,3}(\MGUT)=\hat g_{\smallGUT}$ is missed in \HSS.  The
  gray band is obtained by scanning supersymmetric mass parameters by
  up to a factor 3 above or below the scale $\tm$, under the condition
  of reproducing the observed Higgs mass.
  \label{fig:GUT}}
\end{figure}

Next, we investigate how supersymmetric threshold corrections affect
the GUT condition of gauge-coupling unification in \HSS. We employ our
full one-loop threshold corrections to the MSSM couplings $\hat g_1,
\hat g_2, \hat g_3$ and $\hat y_t$ in order to compute the values of
these couplings at the matching scale $\tm$. The couplings are then
evolved to high energy using the two-loop RGE of the MSSM.  In
figure\fig{GUT} we show the minimum amount (in percent) by which one
coupling $\hat g_i(\MGUT)$ should be changed in order to achieve an
exact crossing $\hat g_1(\MGUT)=\hat g_2(\MGUT)=\hat g_3(\MGUT)$ at
some $\MGUT$, neglecting GUT-scale thresholds.  The gray band is
obtained by scanning the SUSY mass parameters by up to a factor 3
above or below the scale $\tm$, and $A_t$ within the range allowed by
vacuum stability, with $\tan\beta$ adjusted so as to reproduce the
measured value of the Higgs mass. For comparison, in the SM
$g_2(\MGUT)$ is larger than the value corresponding to perfect
unification by approximately $3.5\%$. The figure shows that in \HSS\
perfect gauge-coupling unification can still be achieved as long as
the SUSY scale $\tm$ is lower than a few times $10^6$~GeV.

\subsubsection*{Tuning of the EW scale}

\begin{figure}[t]
$$\includegraphics[width=0.45\textwidth]{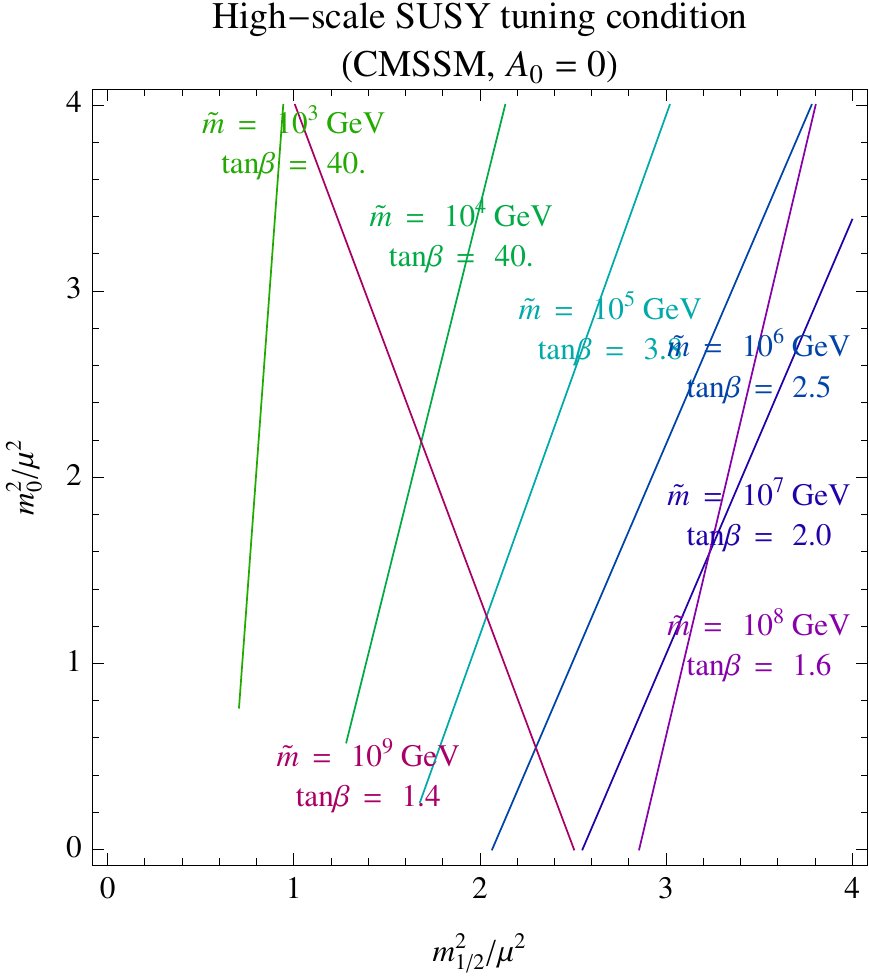}\qquad
\includegraphics[width=0.45\textwidth]{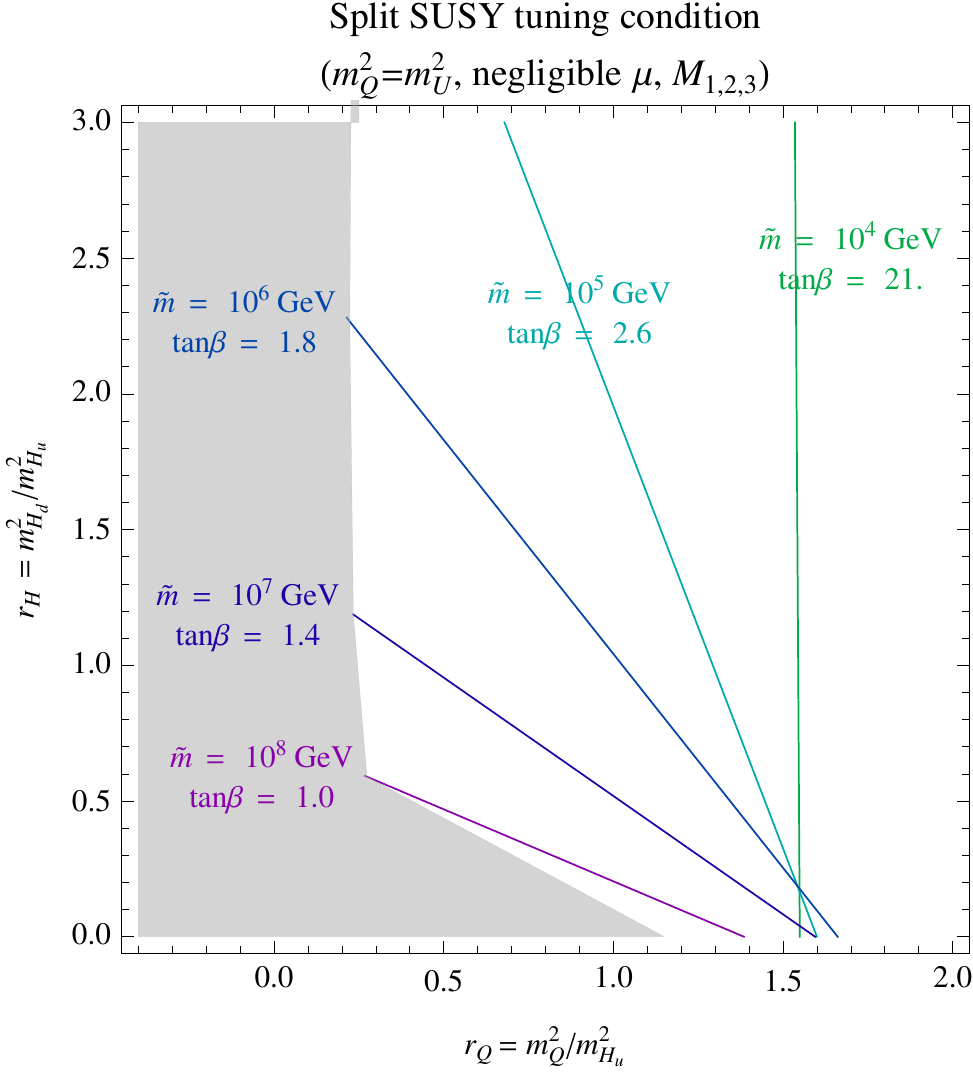}$$
\caption{\em {\bf Left:} Prediction of the SUSY-breaking scale $\tm$
  and the value of $\tan\beta$ from the EW tuning condition and the
  Higgs mass, in \HSS\ with universal gaugino mass $\M5$ and scalar
  mass $m_0$ at the GUT scale (with $A_0=0$). The prediction is
  plotted as a function of the ratios $\M5^2/\mu^2$ and $m_0^2/\mu^2$
  evaluated at the GUT scale. The lines are truncated when the
  vacuum-stability condition is violated. {\bf Right:} same as in the
  left plot, in \sps\ with SU(5) relations for the scalar masses.  The
  prediction is plotted as a function of the ratios $m_Q^2/m_{H_u}^2$
  and $m_{H_d}^2/m_{H_u}^2$ evaluated at the GUT scale. In the shaded region, 
  the EW vacuum is unstable.
  \label{fig:rQrH}}
\end{figure}

The measurement of the Higgs mass has been a crucial new element for
all schemes of \HSS\ because it provides direct information (although
blurred by the unknown parameter $\tan\beta$) on the SUSY-breaking
scale $\tm$. Moreover, although such unnatural schemes do not provide
any dynamical explanation for the tuning of the EW scale, the very
existence of the tuning condition
\beq
\tan^2\beta =\left. \frac{m_{H_d}^2+\mu^2}{m_{H_u}^2+\mu^2}\right|_{\tm} 
\label{sutun}
\eeq
can teach us something about the pattern of SUSY breaking at energy
scales much higher than $\tm$.

Let us consider a simple pattern of SUSY breaking, in which the
superparticle masses at the GUT scale are described by a common
gaugino mass $\M5$, a common scalar mass $m_0$ and a higgsino mass
$\mu$ (for simplicity we set the trilinear coupling $A_0 =0$). For any
given value of the ratios $\M5/\mu$ and $m_0/\mu$, the measured value
of the Higgs mass and the EW tuning condition in \eq{sutun} determine
both $\tan\beta$ and the overall scale of supersymmetry breaking,
$\tm$. We show this prediction in the left plot of
figure~\ref{fig:rQrH}, taking into account the constraint that the
vacuum does not break color spontaneously. The fact that solutions are
found in most of the plane illustrates the effectiveness of radiative
EWSB in supersymmetry. We also note that the quasi-natural solution
($\tm = 10^3$~GeV) with large $\tan\beta$ corresponds to a nearly
vertical line. This is the well-known focus-point behavior,
characteristic of low-energy SUSY with universal boundary
conditions. The novel result is that the model has a second focus
point (in which the tuning condition is approximately independent of
$m_0$) at $\tm = \hbox{few}\times 10^8\GeV$ and low $\tan\beta$.

\subsection{Split SUSY}
\label{sec:split}

Another interesting (albeit unnatural) pattern of SUSY breaking is
given by \sps~\cite{split1,split2,split3} (see
also~\cite{Wells:2003tf,Wells:2004di}). The original idea employs two
independent mass scales. Scalar masses and $B_\mu$ (the mass mixing
between the two scalar components of the Higgs superfields) --- which
correspond to dimension-two, $R$-neutral operators induced by an
effective $D$-term supersymmetry breaking --- are characterized by the
mass parameter $\tm$. Gaugino/higgsino masses and trilinear couplings
$A$ --- which correspond to dimension-three, $R$-charged operators
induced by an effective $F$-term supersymmetry breaking --- are
assumed to be around the weak scale. This spectrum separation can be
naturally justified by the different operator dimensionality, by an
approximate $R$-symmetry, or by the pattern of supersymmetry
breaking. On the other hand, the smallness of the Higgs vacuum
expectation value requires a fine-tuning of order $v^2/{\tm}^2$.

We update here the analysis of the Higgs mass in \sps\ presented in
ref.~\cite{Giudice:2011cg}, by including our improved calculation of
the matching conditions at the scale $\tm$.  The results are shown in
the plots on the right of figures\fig{HeavySUSY}
and\fig{deltalambdaSUSY}, which are the \spsh\ analogs of the
already-described \HSS\ plots on the left of the same figures.

The right plot in figure\fig{HeavySUSY} shows $\MH$ as function of the
common mass $\tm$ of a degenerate scalar spectrum.  We assume that
gauginos and higgsinos are also degenerate with masses
$M_1=M_2=M_3=\mu=$~1~TeV, and we show only lines corresponding to
$A_t=0$ (since in \sps\ $A_t/\tm \ll 1$).

The right plot in figure\fig{deltalambdaSUSY} shows the allowed region
in the ($\tm,\tan\beta$) plane.  The solid black curve shows again the
result in the mass-degenerate case described above.  The light-red
band shows the broadening of the prediction as the scalar mass
parameters are varied between $\tm /3$ and $3\, \tm$.  Finally, the
dark-red band shows the broadening of the prediction of \sps\ (with
universal scalar masses) as $M_t$ is varied in its $1\sigma$ range.

The smallness of $A_t$ and $\mu$ in \sps\ implies that the stop
threshold corrections to the Higgs quartic coupling are smaller than
in the case of \HSS. For this reason, in figure\fig{deltalambdaSUSY}
the light-red band is narrower than the light-blue band, for any given
$\tan\beta$. Note also that in the case of \sps\ the uncertainty in
$M_t$ affects the extraction of $\tm$ at low $\tan\beta$ less than in
\HSS\ (i.e., the dark-red band is narrower than the dark-blue band
near $\tan\beta =1$). The reason is that the Higgs-mass prediction in
\sps\ (unlike the case of \HSS) rises markedly with $\tm$ in the
region around $\MH = 125$~GeV.

The LHC data on the Higgs mass rule out the case of large $\tm$. In
particular, $\tm$ is below $100$~TeV if $\tan\beta >4$ and, for
$\tan\beta$ close to 1, it can only reach about $10^8$~GeV (see
figure\fig{deltalambdaSUSY}).\footnote{See ref.~\cite{Benakli:2013msa}
  for a modified \spsh\ scenario in which $\tm$ can be raised all the
  way up to the GUT scale.}  The determination of $\tm$ can be
translated into a prediction for the gluino
lifetime~\cite{Gambino:2005eh}
\beq c\tau_{\tilde g} = \left(
\frac{2~{\rm TeV}}{M_{\tilde g}}\right)^5\left( \frac{\tilde
  m}{10^7~{\rm GeV}}\right)^4 ~0.4~{\rm m}.  
\eeq 
The Higgs-mass constraint still allows for a wide variety of gluino
decay lengths at the LHC. The mean gluino decay length can be larger
than the size of the detectors ($c\tau_{\tilde g}\simgt 10$~m) for
$\tan\beta$ very close to 1, it is observable as a displaced vertex
($c\tau_{\tilde g}\simgt 50\, \mu$m) for larger $\tan\beta$, while for
$\tan\beta \simgt 2$ the gluino decays promptly.

It is particularly interesting to consider the implications of the EW
tuning condition in the case of
\sps~\cite{Ibarra:2005vb,Delgado:2005ek}, because the theory at $\tm$
contains fewer parameters than \HSS. The EW tuning condition in
\eq{sutun} now becomes
\beq
\tan^2\beta =\left. \frac{m_{H_d}^2}{m_{H_u}^2}\right|_{\tm} \ .
\label{tuning}
\eeq
In the simplified case in which the soft scalar masses satisfy SU(5)
unification relations at the GUT scale $\MGUT$, \eq{tuning} can be
expressed in terms of the two ratios of masses
\beq
r_H = \left. \frac{m_{H_d}^2}{m_{H_u}^2}\right|_{\MGUT} ~,~~~~~
r_Q = \left. \frac{m_{Q}^2}{m_{H_u}^2}\right|_{\MGUT} ~.
\eeq
Here $m_{Q}$ denotes the masses of the left and right stop states,
which belong to the same irreducible representation of SU(5) and thus
are equal at the GUT scale. The EW tuning condition can be expressed
as~\cite{Delgado:2005ek}
\beq
\tan^2\beta = \frac{{\hat K} + \omega +2{\hat K}r_Q 
+(1-\omega )r_H}{1-{\hat K} - \omega- 2{\hat K}r_Q +\omega r_H} \ ,
\label{tune1}
\eeq
where
\beq
{\hat K}=\frac{\sin^2\beta}{2}\left[ 1-\frac{g_t^2(\tm)}{g_t^2(\MGUT)}
\left( \frac{g_3^2(\tm)}{g_3^2(\MGUT)}\right)^{-\frac{16}{9}}
\left( \frac{g_2^2(\tm)}{g_2^2(\MGUT)}\right)^{3}
\left( \frac{g_1^2(\tm)}{g_1^2(\MGUT)}\right)^{\frac{13}{99}}\right] \ ,
\eeq
\beq
\omega = \frac{1}{22} \left[ 1- \frac{g_1^2(\tm)}{g_1^2(\MGUT)}\right] \ .
\eeq
Note that we have defined $\hat K$ in such a way that it is
independent of $\tan\beta$ at the leading order. For $\tm=\MGUT$ we
have $\hat K =\omega =0$, while for $\tm=10$~TeV we find $\hat K =
0.28$ and $\omega =0.024$.

Furthermore, the conditions $m_Q^2,m_U^2>0$ (no color-breaking minima)
and $m_{H_u}^2+m_{H_d}^2>0$ (stability of the EW vacuum) imply the
restriction
\beq
-\frac{\omega}{1-\omega}<r_H< \frac{ (4-9\omega )\tan^4\beta 
+  (7-4{\hat K}\omega -9\omega )\tan^2\beta 
-4{\hat K}\omega}{3\omega  (1-4{\hat K})\tan^4\beta 
+ [-3(1-\omega ) +{\hat K} (1-4\omega )] \tan^2\beta +4(1-\omega ) {\hat K}}
 \ .
\label{tune2}
\eeq

Given the values of $r_H$ and $r_Q$, the theory predicts both $\tilde
m$ and $\tan\beta$ from the Higgs mass and the EWSB condition. Our
results are shown in the right plot of figure~\ref{fig:rQrH}. It is
remarkable that acceptable solutions are found in a region close to
universality, where both $r_H$ and $r_Q$ are of order one.  A further
restriction corresponds to the case $r_H=1$, in which there is
unification of the Higgs mass parameters, $m_{H_d}^2=m_{H_u}^2$, at
the GUT scale. Such equality is not uncommon in models arising from
compactified string theory~\cite{Ibanez:2013gf}. Then, the EW tuning
conditions in eqs.~(\ref{tune1}) and (\ref{tune2}) become
\beq
r_Q =\frac{(1-{\hat K})\tan^{2}\beta -1-{\hat K}}{2{\hat K} 
(1+\tan^{2}\beta )}~,~~~~
\tan^2\beta > \frac{3-4{\hat K}(1+\tan^{-2}\beta )}{3-7{\hat K} 
(1+\tan^{-2}\beta )}~.
\label{higuni}
\eeq
The result can be read from the right plot in figure~\ref{fig:rQrH}
along the horizontal line $r_H=1$. The case of exact universality
corresponds to the point with $r_H=r_Q=1$. The EW tuning condition now
becomes
\beq
\tan^2\beta =\frac{1+3{\hat K}}{1-3{\hat K}} \ .
\label{eqtanb}
\eeq
The prediction for $\tan\beta$ from EW tuning with universal scalars
at the GUT scale in \eq{eqtanb} is shown as a dot-dashed line in the
right plot of figure\fig{deltalambdaSUSY}.  We find that exact
universality corresponds to $\tm\approx 10^6\GeV$ and $\tan\beta \approx2$, 
for central values of the SM parameters.

\section{Mini-split with anomaly mediation}\label{sec:sms}

The particular range of values of $\tm$ determined by the Higgs-mass
measurement have fueled interest in a simple version of \sps\ emerging
from anomaly mediation~\cite{Randall:1998uk,Giudice:1998xp}. The model
was originally proposed in ref.~\cite{Giudice:1998xp} and its spectrum
was reconsidered in
refs.~\cite{Wells:2003tf,Wells:2004di,split1,split2,split3}. In
ref.~\cite{welltemp} it was recognized as the simplest model of
\sps\ and its connection with dark matter was elucidated. Subsequent
studies are contained in
refs.~\cite{Hall:2011jd,Kane:2011kj,Acharya:2012tw,Bhattacherjee:2012ed,
  Arvanitaki:2012ps,Hall:2012zp,ArkaniHamed:2012gw,Moroi:2013sfa,
  McKeen:2013dma,Kahn:2013pfa}.

\medskip

The original motivation of mini-split with anomaly mediation is
linked to the observation that gaugino masses require supersymmetry
breaking through the $R$-charged $F$-term of a chiral superfield $S$
which must be a singlet under all gauge and global charges
\beq
M_{i} ~~~\to ~~~\int d^2\theta \frac{S}{M_*}W_{i\alpha} W_i^\alpha ~.
\label{mgluin}
\eeq
Here $W_{i\alpha}$ ($i=1,2,3$) is the gauge-strength chiral
superfield, and $M_*$ is the mediation scale. On the other hand, masses
for the scalar components of chiral superfields $Q$ are induced by the
$F$-term of any chiral superfield $X$, irrespectively of its global,
gauge, or $R$ charges,
\beq     
{\tm}^2_{Q} ~~~\to ~~~\int d^4\theta \frac{X^\dagger X}{M_*^2}~Q^\dagger Q ~.
\eeq
This difference becomes important especially in models with dynamical
supersymmetry breaking, where no singlets are present. In this case,
while scalars acquire a tree-level mass $\tm =F_X/M_*$ (where
$\sqrt{F_X}$ is the scale of supersymmetry breaking, $X=\theta^2
F_X$), the leading contribution to a gaugino mass $M_i$ comes
from one-loop anomaly mediation effects
\beq 
\label{gauginomass}
M_{i} =\frac{\beta_i}{g_i}\,m_{3/2} \, ,
\eeq
where $g_i$ is the corresponding gauge coupling, $\beta_i$ its beta
function, and $m_{3/2}$ is the gravitino mass.

The same reasoning leads us to conclude that also $A$-terms receive
their main contribution from anomaly mediation
\beq
A_y =-\frac{\beta_y}{y}\,m_{3/2} \, ,
\eeq
where $y$ is the corresponding Yukawa coupling and $\beta_y$ its beta
function.

\smallskip

In order to complete the setup, we have to specify how $\mu$ and
$B_\mu$ are generated. In general, we expect tree-level contributions
to $B_\mu$ of order ${\tm}^2$ induced by
\beq     
\int d^4\theta \,\frac{X^\dagger X}{M_*^2}~H_uH_d ~,
\label{bbbmu}
\eeq
while $\mu$ is not generated at this level. The most interesting (and
plausible) possibility is that $\mu$ is generated by the same
mechanism that gives mass to gauginos:
gravity~\cite{Giudice:1988yz}. The difference is that, while gaugino
masses are generated at one loop, gravity induces the $\mu$ term at
tree level. This can be seen by writing the relevant supergravity
terms involving the Higgs superfields ${\hat H}_{u,d}$ (normalized
such that they have zero canonical dimension) and the conformal
compensator $\Phi$, with canonical dimension $d_\Phi=1$,
\beq
\int d^4\theta\, \Phi^\dagger \Phi \left[ {\hat H}_{u,d}^\dagger 
{\hat H}_{u,d} +\left( c\, {\hat H}_{u}{\hat H}_{d}  
+{\rm h.c.}\right) \right] \, .
\label{higla}
\eeq
In terms of the canonical superfields $H_{u,d}=\Phi {\hat H}_{u,d}$,
the lagrangian in \eqg{higla} becomes
\beq
\int d^4\theta \left[ {H}_{u,d}^\dagger {H}_{u,d} +\left( c\, 
\frac{\Phi^\dagger}{\Phi}{H}_{u}{H}_{d} +{\rm h.c.}\right) \right] \, .
\eeq
While usually, because of scale invariance, the dependence on $\Phi$
drops from the classical lagrangian once the latter is expressed in
terms of canonically-normalized fields and it reappears only through
the scale anomaly in the $\beta$-functions, here we have an explicit
dependence on $\Phi$ at tree level. After supersymmetry breaking
($\Phi \propto 1- m_{3/2}\,\theta^2$), we find
\beq
\mu =c\, m_{3/2},~~~~B_\mu =c\, m_{3/2}^2.
\label{mbmu}
\eeq
If the coupling constant $c$ is of order one, then the typical mass
scale of both $\mu$ and $B_\mu$ is the gravitino mass $m_{3/2}$, which
is parametrically equal to the scalar-mass scale $\tm$ when the
mediation scale $M_*$ is close to the Planck scale. However, $c$ is
the only PQ-breaking parameter in the theory and it could be small for
symmetry reasons. Thus,
$\mu$ could take any value between $m_{3/2}$ and the weak
scale. Irrespectively of the value of $c$, \eqg{mbmu} implies
\beq
B_\mu =\mu\, m_{3/2} \, .
\label{mubmu}
\eeq
This is an interesting relation because it allows us to link the value
of $\mu$ directly to $\tan\beta$. Indeed, after EW breaking, we have
$\sin2\beta =2|B_\mu| /\mA^2$, where $\mA$ is the Higgs pseudoscalar
mass, and thus
\beq
\sin2\beta =\frac{2|\mu| m_{3/2}}{\mA^2}\, .
\label{tanbb}
\eeq
The expression of the pseudoscalar mass
$\mA^2=m_{H_u}^2+m_{H_d}^2+2\mu^2$ can be simplified with the help of
the EW tuning condition \eq{sutun} and becomes
\beq
\mA^2 = (1+\tan^{-2}\beta ) (m_{H_d}^2+\mu^2) \, .
\label{pseudom}
\eeq
Hence, \eq{tanbb} can be rewritten as
\beq
\tan\beta =  \frac{m_{H_d}^2}{|\mu|\ m_{3/2}}+\frac{|\mu|}{m_{3/2}}\, .
\label{tanbb2}
\eeq

\bigskip

We can now summarize the features of the spectrum of mini-split with
anomaly mediation.

\begin{itemize}
\item[{\it (i)}] {\bf Squarks and sleptons:} Supersymmetric scalars are
  characterized by the mass scale $\tm$, although the details of the
  spectrum are not calculable. The typical size of scalar mass is
  related to the gravitino mass by $\tm \approx (\mpl /M_*)\,m_{3/2}$,
  where $M_*$ is the mediation scale.  The requirement that $M_*$ is
  larger than the unification scale, in order not to affect gauge
  coupling unification, implies $1\simlt \tm /m_{3/2} \simlt
  10^3$. However, the simplest possibility is that $\tm /m_{3/2}$ is
  of order one and gravity is the only mediator of supersymmetry
  breaking.

\item[{\it (ii)}] {\bf Gauginos:} Anomaly mediation gives precise
  predictions for the physical masses of the gauginos, in terms of
  $m_{3/2}$~\cite{Gherghetta:1999sw,Gupta:2012gu}. In our analysis we
  include the next-to-leading-order corrections controlled by the
  strong and top-Yukawa couplings, as well as two important effects
  controlled by the weak gauge couplings. Of the latter, the first
  consists of logarithms of ${\tm}/M_i$, which take into account how
  the gaugino masses deviate from the anomaly-mediation trajectory
  after squarks and sleptons are integrated out. The second effect is
  relevant when $\mu$ is larger than the weak scale. In this case, the
  Higgs superfields act as messengers of supersymmetry breaking and
  give a one-loop contribution to the gaugino masses proportional to
  $\mu$. Assuming common mass terms for the squarks and for the
  sleptons, the physical gaugino masses are
 \begin{eqnsystem}{sys:M123}
 \Mbino
&=&M_1(Q)\, 
  \bigg[1+\frac{C_\mu}{11} \nonumber 
   +  \frac{g_1^2}{80\pi^2}\bigg(
-\frac{41}{2}\,\ln\frac{Q^2}{M_1^2}-\frac12\,\ln\frac{\mu^2}{M_1^2}\\
 && 
\qquad 
+\ln\frac{\mA^2}{M_1^2}
+11\,\ln\frac{m^2_{\tilde q}}{M_1^2}
+9\,\ln\frac{m^2_{\tilde \ell}}{M_1^2}
 \bigg) +\frac{g_3^2}{6\pi^2}-\frac{13 g_t^2}{264\pi^2\sin^2\beta} 
     \bigg] , \label{eq:mmbin}\\\nonumber\\ 
\Mwino
&=&M_2(Q)\, 
  \bigg[1+C_\mu +\frac{g_2^2}{16\pi^2}\bigg(
\frac{19}{6}\,\ln\frac{Q^2}{M_2^2} -\frac16\,\ln\frac{\mu^2}{M_2^2} \nonumber \\
&&\qquad
+\frac13\,\ln\frac{\mA^2}{M_2^2}
+3\,\ln\frac{m^2_{\tilde q}}{M_2^2}
+\ln\frac{m^2_{\tilde \ell}}{M_2^2}
 \bigg)   +\frac{3 g_3^2}{2\pi^2}-\frac{3g_t^2}{8\pi^2\sin^2\beta}
    \bigg] ,\label{eq:mmwin}\\ \nonumber\\ 
\Mgino&=&M_3(Q)\,  
\bigg[1+\frac{g_3^2}{16\pi^2}\bigg( 
7\,\ln\frac{Q^2}{M_3^2}
+4\,\ln\frac{m^2_{\tilde q}}{M_3^2}
+\frac{25}{3}
-2\, F\big(\frac{M_3^2}{m^2_{\tilde q}}\big) 
\bigg)+\frac{g_t^2}{12\pi^2\sin^2\beta}\bigg] \, ,\nonumber\\ 
\end{eqnsystem}
where
\beq
M_1(Q)=\frac{33\,g_1^2(Q)}{80\pi^2}\, m_{3/2}\, ,\qquad
M_2(Q) =  \frac{g_2^2(Q)}{16\pi^2} \, m_{3/2}\, ,\qquad
M_3(Q) = -\frac{3\,g_3^2(Q)}{16\pi^2} \, m_{3/2}\,,\eeq
$g_i(Q)$ are the gauge couplings of the SM renormalized in the
$\msbar$ scheme at a generic scale $Q$, and
\beq
\label{eq:Cmu}
C_\mu = \frac{\mu}{m_{3/2}} \, \frac{\mA^2\sin2\beta }{\mA^2-\mu^2}\, \ln
\frac{\mA^2}{\mu^2} \ , 
\eeq 
\beq
F(x) ~=~ 3\left[ \frac32-\frac1x 
-\left(\frac1x -1\right)^2\,\ln|1-x|\right] ~=~x+{\cal O}(x^2)~.
\eeq

\item[{\it (iii)}] {\bf Higgsinos and ${\mathbf \tan\beta}$:} The
  higgsino mass $\mu$ is expected to be of order $m_{3/2}$, if there
  is no suppression related to PQ breaking. Otherwise, $\mu$ is a free
  parameter, which could vary between $m_{3/2}$ and the weak
    scale. In general, $B_\mu$ is of order ${\tm}^2$ and $\tan\beta$
  could take any value. However, when $\mu$ and $B_\mu$ are generated
  by the same operator and \eqg{mubmu} holds, then $\tan\beta$ is
  determined according to \eqg{tanbb2}.
\end{itemize}

Mini-split with anomaly mediation has several theoretical and
phenomenological attractive features. It retains the positive aspects
of \sps\ (gauge coupling unification, dark matter candidates, easing
of the flavor problem) without requiring the artificial (although
possible~\cite{split1,split3,Luty:2002ff}) suppression of one-loop
anomaly-mediated gravitational contributions. It retains the positive
aspects of anomaly mediation (elegance, predictivity, viability of
dynamical supersymmetry breaking) without introducing the problem of
tachyonic sleptons~\cite{Randall:1998uk}. Moreover, the most relevant
point for our present analysis is that mini-split with anomaly
mediation gives a prediction for the Higgs mass in the right range, as
we will show in the next section. Of course the drawback is that the
theory is not technically natural.

\subsection{Phenomenology of mini-split with anomaly mediation}

The theory is essentially described in terms of 4 parameters: $\tm$,
$m_{3/2}$, $\mu$ and $\tan\beta$. One of these parameters is fixed by
the value of the Higgs mass, and two more parameters can be fixed
under certain assumptions. If gravity is the only mediator of
supersymmetry breaking, $\tm$ is roughly equal to $m_{3/2}$. If we
assume that the operator in \eq{bbbmu} is absent and that both $B_\mu$
and $\mu$ originate from the operator in \eq{higla}, then the value of
$\tan\beta$ is given by \eq{tanbb2} with $m_{H_d}^2=\tm^2$.

\begin{figure}[t]
$$
\includegraphics[height=8cm]{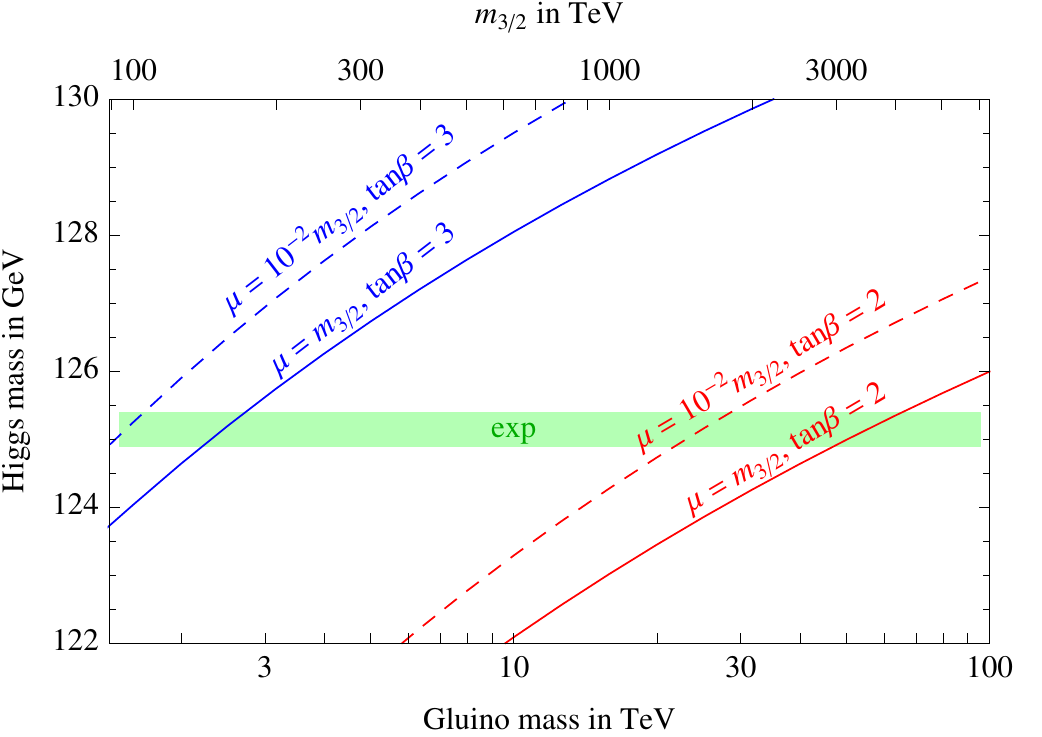}
$$
\caption{\em Predicted Higgs mass in mini-split with anomaly mediation, 
 in the case of soft scalar masses equal to $m_{3/2}$. The result is shown 
 as a function of the gluino mass $\Mgino$ (or, equivalently, the gravitino 
 mass $m_{3/2}$) for different values of $\tan\beta$ and of $\mu$.
\label{fig:MiniSplitMh}}
\end{figure}

The scale of supersymmetry breaking in mini-split with anomaly
mediation is very favorable for explaining the observed Higgs mass. As
an example, figure\fig{MiniSplitMh} shows the prediction of the Higgs
mass as a function of the gluino mass $\Mgino$ (or, equivalently, of
the gravitino mass $m_{3/2}$), in the special case in which all soft
scalar masses are equal to $m_{3/2}$. A successful prediction is
obtained for $\tan\beta$ in the range between 2 and 3. We also show
the impact of varying $\mu$ between $m_{3/2}$ and the gaugino mass
scale: the change in the Higgs mass is mild, in the range of 2--3~GeV.

In spite of having relatively few free parameters, the theory has a
rich variety of possibilities for the nature of the LSP and this has
important implications for dark matter. While most studies focused on
the case in which $\mu$ is of the order of the gaugino masses, new
possibilities for DM appear when $\mu$ is allowed to vary (for some
studies of the case $\mu ={\cal O}(m_{3/2})$, see
refs.~\cite{ArkaniHamed:2012gw,Harigaya:2014dwa}). The important
parameter that defines the nature of the LSP is $C_\mu$, which is
defined in \eq{eq:Cmu} and describes the source of electroweak gaugino
masses coming from the breaking of supersymmetry in the Higgs-higgsino
system.  In the ordinary case of anomaly mediation with $\mu$ at the
weak scale, $C_\mu$ is ${\cal O} (\alpha / 4 \pi)$, thus its
contribution to gaugino masses is parametrically equal to the one-loop
corrections in eqs.~(\ref{eq:mmbin}) and (\ref{eq:mmwin}). However,
when $\mu$ is of the same size as $m_{3/2}$, the parameter $C_\mu$ is
of order unity and its contribution to gaugino masses is comparable to
the leading effect in anomaly mediation. In the special case in which
both $B_\mu$ and $\mu$ are generated by \eq{higla}, $C_\mu$ becomes a
function of a single mass ratio,
\beq |C_\mu | = \frac{2 \ln(
  {\mA^2}/{\mu^2})}{{\mA^2}/{\mu^2}-1} \ .  
\eeq

\begin{figure}[t]
$$
\includegraphics[height=8cm]{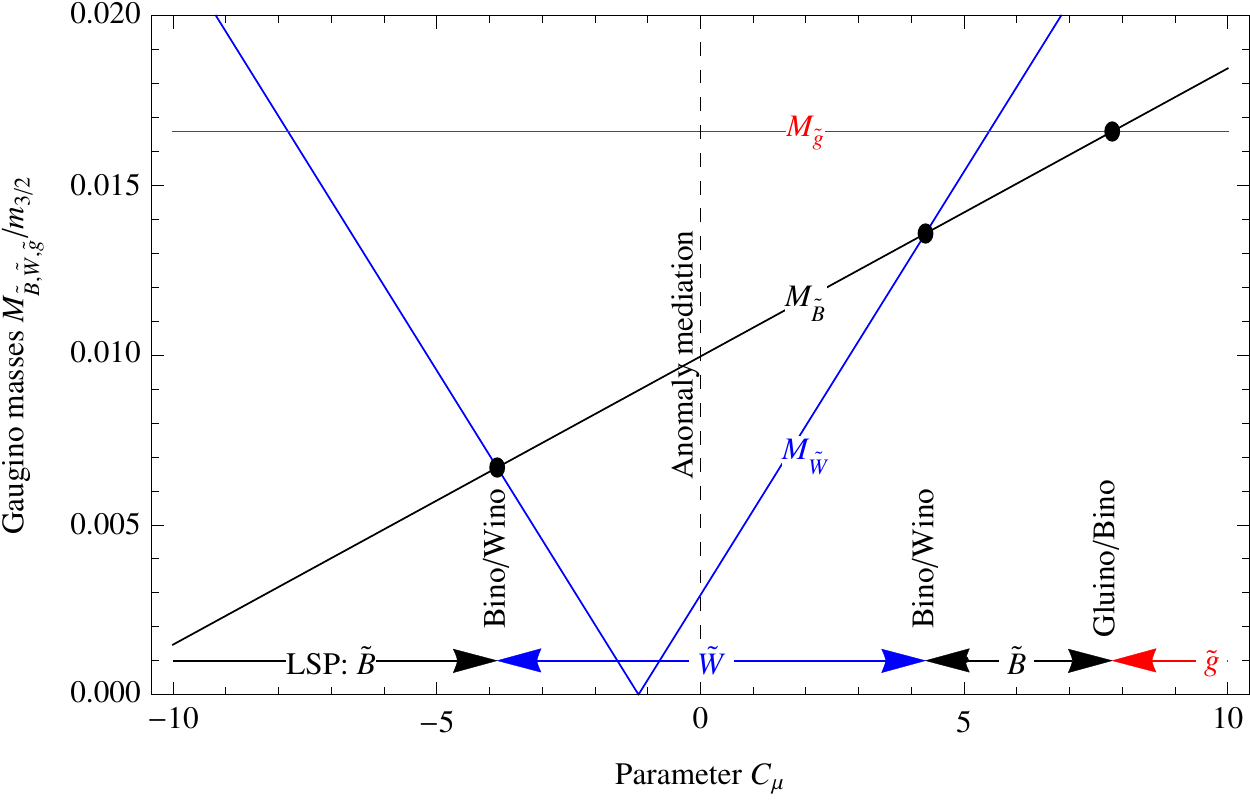}
$$
\caption{\em Physical gaugino masses in units of $m_{3/2}$ in
  mini-split with anomaly mediation, as a function of the parameter
  $C_\mu$ defined in \eq{eq:Cmu}.  
\label{fig:models2}}
\end{figure}

Depending on the value of $C_\mu$, the gaugino mass spectrum and the
nature of the LSP change, as illustrated in figure~\ref{fig:models2},
where the three gaugino masses (in units of $m_{3/2}$) are plotted as
functions of $C_\mu$. This change in the mass spectrum is important
for two reasons. First, different options for the LSP allow for a
richer variety of DM candidates with different perspectives for
discovery in DM detection experiments. Second, a more compressed
gaugino spectrum increases the chances of discovery at the LHC, once
the overall mass scale is fixed by DM relic abundance arguments. The
various alternatives for the LSP are the following.

\begin{itemize}
\item {\bf Wino LSP:} For $|C_\mu|\circa{<}4$, the LSP is the Wino. 
This case
  includes the usual mass spectrum of \sps\ with $\mu$ at the EW
  scale. If the Wino is a DM thermal relic, then $\Mwino=2.7$~TeV.
  The model is outside the reach of the LHC, even in the most
  favorable case in which $C_\mu$ is in its upper range, and
  $\Mgino/\Mwino$ is as small as 1.2. Direct detection of thermal Wino
  DM is difficult, but the prospects from indirect searches are much
  more promising. Current bounds from gamma rays are already rather
  constraining~\cite{Cohen:2013ama,Fan:2013faa}, although very
  dependent on the assumptions on the halo profiles.

\item {\bf Higgsino LSP:} For $C_\mu \approx 0$, the higgsino can be
  the LSP. Thermal relic DM is obtained for a higgsino mass of
  $1.1\TeV$.  This implies the lower bound $\Mgino > 6.6\TeV$. Thus,
  the gluino is too heavy to be probed at the LHC, in the case of a
  thermal relic pure higgsino.

\item {\bf Bino LSP:} For $C_\mu < -3.9$ and $4.1<C_\mu < 7.8$, the
  Bino is the LSP. Its thermal relic abundance would overclose the
  universe, so Bino DM requires some source of late entropy injection
  or low reheat temperature~\cite{Giudice:2000ex,Gelmini:2006pw}.  In
  the window $4.1<C_\mu < 7.8$, the gaugino mass spectrum is fairly
  compressed, with the gluino mass larger than the LSP mass by 20\% or
  less.

\item {\bf Gluino LSP:} For $C_\mu > 7.8$, the gluino is the LSP. This
  case is not acceptable for DM, but it could be interesting for
  collider searches. The gluino could escape cosmological constraints
  with the help of small $R$-violating effective interactions that make
  the LSP unstable. From the collider point of view, the gluino can
  behave as a stable, unstable, or long-lived particle, depending on
  the strength of the effective $R$-violation.

\item {\bf Bino-Wino LSP:} For $|C_\mu | \approx 4$, the LSP can be a
  well-tempered Bino-Wino. For 10\% mass splittings, the mass of the
  LSP can be in the range of several hundred GeV~\cite{welltemp}. We
  find that $\Mgino/\Mwino=2.4$ (for $C_\mu \approx -4$) and
  $\Mgino/\Mwino=1.2$ (for $C_\mu \approx 4$). So these cases are
  particularly favorable for the LHC: the DM particle can be
  light, the gluino is not much heavier than the LSP, and their mass
  ratio is precisely determined.

\item {\bf Higgsino-Wino LSP:} For $C_\mu \approx 0$, the LSP can be a
  mixture of higgsino and Wino. Not much is gained in terms of relic
  abundance, since both the higgsino and the Wino have relatively
  large annihilation cross sections, but the detection rate in direct
  DM experiments can be sizable due to Higgs-boson exchange.

\item {\bf Gluino-Bino LSP:} The value $C_\mu \approx 7.8$ allows for
  the unusual possibility of coannihilation between gluino and
  Bino. This case was recently discussed in
  ref.~\cite{deSimone:2014pda}. For mass splittings in the 100--150
  GeV range, the Bino can be a thermal relic DM and the gluino be
  within reach of the LHC. However, the experimental search for
  gluinos is made difficult by their soft decay products. At present,
  the LHC bound on the gluino mass completely evaporates as soon as
  the LSP mass is larger than
  500--600~GeV~\cite{Chatrchyan:2014lfa,Aad:2014wea}.
\end{itemize}

\begin{figure}[t]
$$
\includegraphics[width=0.45\textwidth]{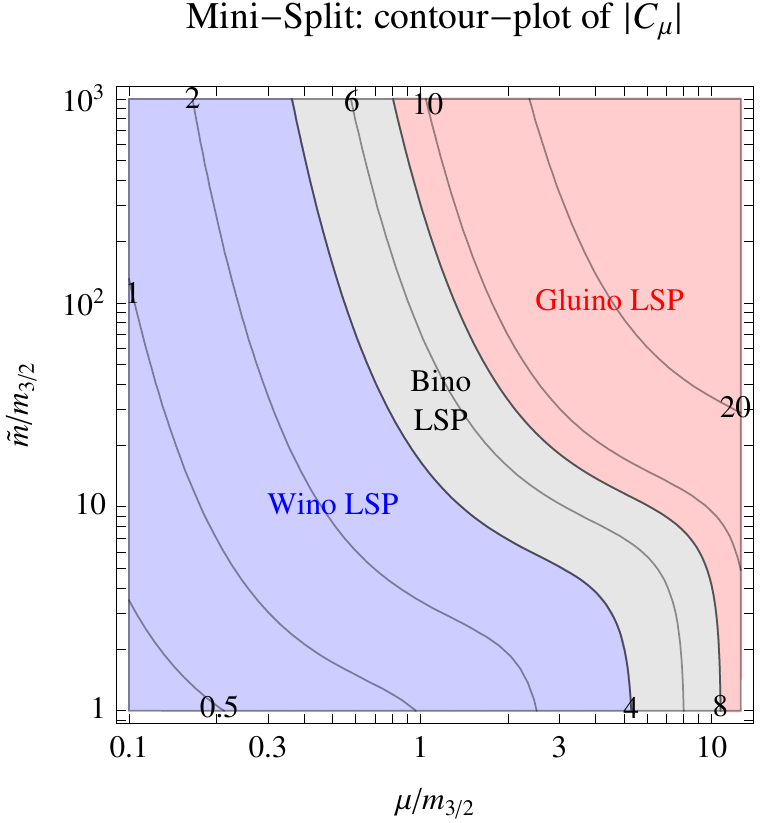}\qquad
\includegraphics[width=0.45\textwidth]{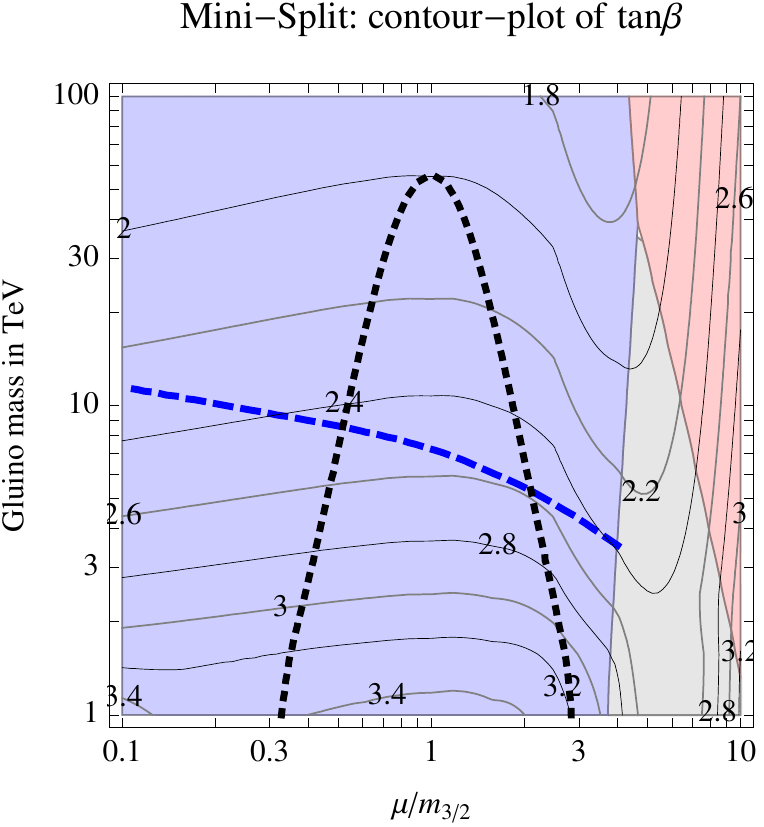}$$
\caption{\em {\bf Left:} $C_\mu$ as a function of the mass ratios $\mu
  /m_{3/2}$ and $\tm /m_{3/2}$. {\bf Right:} the value of $\tan\beta$
  that reproduces the Higgs mass as a function of $\mu/m_{3/2}$ and of
  the physical gluino mass $\Mgino$ for $\tm = m_{3/2}$.  Along the
  dashed blue curve, the Wino thermal DM abundance reproduces the
  observed DM density.  Along the short-dashed black curve $\tan\beta$ as
  predicted by \eq{tanbb2} reproduces the observed $\MH$.  In the blue
  region, the LSP is the Wino; in the gray region, the LSP is the
  Bino; in the red region, the LSP can be either the gluino or the
  Bino, depending on the sign of $\mu$; Higgsino DM is obtained for
  $|\mu/m_{3/2}|\circa{<} 0.003 $.
  \label{fig:models}}
\end{figure}

It is useful to express $C_\mu$ in terms of the original parameters of
the model. Using the expression of the pseudoscalar mass in
\eq{pseudom}, we can rewrite \eq{eq:Cmu} as
\beq
C_\mu = \frac{2\mu \tan\beta}{m_{3/2}}\ 
\frac{\tm^2+\mu^2}{(\tan^2\beta +1)\tm^2+\mu^2}\ 
\ln \left[(1+\tan^{-2}\beta)\left( 1+\frac{\tm^2}{\mu^2}\right)\right] \, .
\eeq
In the left panel of figure\fig{models} we plot $C_\mu$ as a function
of the mass ratios $\mu /m_{3/2}$ and $\tm /m_{3/2}$, fixing
$\tan\beta$ with the requirement of a correct value for the Higgs
mass.  This shows that values of $|C_\mu|$ in the range 1--10 can be
easily obtained for natural choices of the fundamental parameters.
Finally, in the right panel of figure~\ref{fig:models} we present the
map of the various LSP regions in a plane spanned by the physical
gluino mass and $\mu /m_{3/2}$, under the restrictive assumption of
exact universality of scalar masses with $\tm =m_{3/2}$. In this plane
we show contours of $\tan\beta$, extracted from the Higgs mass
measurement. This figure illustrates once again the mild dependence on
$\mu$ of the prediction for $\tan\beta$ (or, equivalently, for the
Higgs mass), as $\mu$ is varied from $m_{3/2}$ to the gaugino
  mass scale. The assumption of \eq{tanbb2} (with
$m_{H_d}^2=\tm^2=m_{3/2}^2$) fixes one extra parameter and constrains
the theory to live along the black short-dashed line.

\section{Conclusions}

As collider and DM experiments keep on setting more stringent
constraints on low-energy SUSY, the interest is shifting towards
models in which supersymmetry is broken at a scale larger than the
natural scale $M_Z$. In this paper, we have performed a thorough
analysis of the Higgs mass in such ``unnatural" models.

Our main new computational result is contained in
section~\ref{sec:thre}, where we give complete expressions for the
one-loop threshold corrections to the Higgs quartic coupling
$\lambda$, the top Yukawa coupling, gauge couplings, and gaugino
couplings (for \sps) evaluated at the SM/MSSM, SM/\spsh\ and
\spsh/MSSM matching scales.  For the Higgs quartic coupling $\lambda$,
we include also two-loop QCD threshold corrections. Our results
complete and correct previous literature on the subject.  Furthermore,
we adopt the extraction of SM parameters with NNLO precision, using
the results of ref.~\cite{SMpar}.

\medskip

In section~\ref{sec:hiSUSY}, we applied our results to special
realizations of supersymmetry broken at scales larger than $M_Z$. The
first case refers to {\it quasi-natural SUSY}, in which supersymmetric
particle masses are in the multi-TeV range. Although the scale of
supersymmetry breaking $\tm$ is not far from the electroweak scale, a
precise calculation of the Higgs mass requires resummation of the
logarithms of the ratio $\tm /M_Z$.  Our results are presented in
figure\fig{Mhpred}. We find that the Higgs mass measurement implies
$\tm \circa{>} 10$~TeV (no stop mixing) and $\tm \circa{>} 2$~TeV
(maximal stop mixing), in the case of moderately large $\tan\beta$ and
degenerate supersymmetric mass parameters at the scale $\tm$.

\smallskip

Next, we considered the case of {\it \HSS}, in which we let $\tm$ vary
arbitrarily. Our predictions for the values of $\tm$ and $\tan\beta$
determined by the Higgs mass are shown in figures\fig{HeavySUSY}
and\fig{deltalambdaSUSY}. We used our calculation of the
supersymmetric threshold corrections to show how non-degenerate
spectra affect the Higgs mass determination. In particular, we find
that the Higgs mass measurement implies $\tm \circa{<} 2\times
10^{10}$~GeV (for degenerate supersymmetric particles) and $\tm
\circa{<} 10^{11}\GeV$ (for supersymmetric mass parameters larger or
smaller than $\tm$ by a factor of 3), and for the central value of the
top mass $M_t$.  If $M_t$ is $3\sigma$ lower than its central value,
SUSY up to the Planck scale becomes allowed. Extra Higgs interactions
at large energies can change the picture.  We have also explored the
implications of the relation between $\tm$ and $\tan\beta$ implied by
the measured Higgs mass for gauge-coupling unification and for the
tuning required to generate the low scale of $M_Z$.

\smallskip

We repeated the exercise for {\it \sps}, in which scalar
supersymmetric particles have masses of order $\tm$, while fermionic
supersymmetric particles lie around the weak scale. Our results, shown
in figure\fig{deltalambdaSUSY}, indicate that the Higgs mass
constrains the scale of \sps\ $\tm \circa{<} 6 \times 10^7$~GeV (for
degenerate scalar supersymmetric particles) and $\tm \circa{<}
10^{8}$~GeV (for scalar mass parameters larger or smaller than $\tm$
by a factor of 3).  Decreasing the top mass by $1\sigma$ increases the
maximal $\tm$ by a factor of 2.  For universal scalar mass parameters
at the GUT scale, we find that the Higgs mass and the tuning condition
determine $\tm \approx 10^6\GeV$ and $\tan\beta \approx2$. This
prediction is relaxed as we allow for non-universality of scalar
masses at the GUT scale (see figure\fig{rQrH}). However, the
constraints from EWSB and color conservation select a region of
boundary conditions at the GUT scale centered around complete
universality.

Section \ref{sec:sms} is devoted to the last scenario we considered:
{\it mini-split with anomaly mediation}. In this case, scalar
particles feel supersymmetry breaking at tree level, while gauginos
get mass only from one-loop anomaly mediation effects. We have
discussed various possibilities for the origin of the higgsino mass
$\mu$, which in principle could be anywhere between the gravitino
  and gaugino masses. Changing $\mu$ in this range has a limited
effect on the Higgs mass (see figure\fig{MiniSplitMh}), but a very
important impact on the nature of the LSP. We have found that
mini-split with anomaly mediation, in spite of its few free
parameters, can lead to a variety of possibilities for the LSP and
thus for the DM candidate, as summarized in
figure\fig{models2}. Moreover, it is possible to obtain very
compressed spectra for gaugino masses, in which the gluino does not
lie far beyond the DM particle, thus increasing the chance of LHC
discovery.

\small

\paragraph{Acknowledgments} 
This work was supported by the ESF grant MTT8.
% and by SF0690030s09 project.  E.B.~and P.S.~were supported in part
by the Research Executive Agency (REA) of the European Commission
under the Grant Agreement PITN-GA-2010-264564 (LHCPhenoNet), and by
French state funds managed by the ANR (ANR-11-IDEX-0004-02) in the
context of the ILP LABEX (ANR-10-LABX-63). P.S. thanks the authors of
ref.~\cite{Allanach:2001kg} for useful communication about {\tt
  SoftSusy}, and A.S.\ thanks Alexander Knochel for having pointed out
a mistake in the expressions of~\cite{Giudice:2011cg} concerning the
heavy Higgs loop contribution to the quartic Higgs coupling.

\appendix

\section{Loop functions}\label{F}
The loop functions that describe the stop contribution to the Higgs
quartic coupling are:
\begin{eqnsystem}{sys:Fstop}
\wt F_1(x)&=&\frac{x\ln x^2}{x^2-1}~,\\
\wt F_2(x)&=&\frac{6\,x^2\left[ 2-2\,x^2 
+(1+x^2)\ln x^2 \right]}{(x^2-1)^3}~,\\
\wt F_3(x)&=&\frac{2x [ 5(1-x^2)+(1+4x^2)\ln x^2]}{3(x^2-1)^2}~,\\
\wt F_4(x) &=& \frac{2x ( x^2-1-\ln x^2)}{(x^2-1)^2}~,\\
\wt F_5(x) &=& \frac{3 x (1 - x^4 + 2 x^2 \ln x^2)}{(1 - x^2)^3}~.
\end{eqnsystem}
The extra loop functions that describe the stop contribution to the top Yukawa coupling are:
\begin{eqnsystem}{sys:Filde6}
\wt{F}_6(x) \,&=&\, \frac{x^2-3}{4\,(1-x^2)}
\,+\,\frac{x^2\,(x^2-2)}{2\,(1-x^2)^2}\, \ln x^2\,,\\
\wt{F}_7(x) \, &=&\, \frac{-3 \,(x^4-6 x^2 + 1)}{2\,(x^2-1)^2} 
\,+\, \frac{3 \,x^4 \,(x^2-3)}{(x^2-1)^3}\, \ln x^2\,,\\
\wt{F}_8\left(x_1,x_2 \right)&=&
-2 ~+~\frac{2}{x_1^2-x_2^2}\,\left(
\frac{x_1^4}{x_1^2-1}\ln x_1^2 - \frac{x_2^4}{x_2^2-1}\ln x_2^2\right)~,\\
\nonumber\\
\wt{F}_9\left(x_1,x_2 \right)&=&
\frac{2}{x_1^2-x_2^2}\,\left(
\frac{x_1^2}{x_1^2-1}\ln x_1^2 - \frac{x_2^2}{x_2^2-1}\ln x_2^2\right)~.
\label{moreFs}
\end{eqnsystem}
Finally, the loop functions for the gaugino-higgsino corrections to both 
$\lambda$ and $g_t$ are~\cite{Giudice:2011cg}
\begin{eqnsystem}{sys:f}
f(r)& =& \widetilde F_5(r)~,~~~~~~~~~~~g(r)~=~ \widetilde F_7(r)~,\\\nonumber\\
f_1(r) &=& \frac{6 \left(r^2+3\right) r^2}{7 \left(r^2-1\right)^2}+\frac{6 \left(r^2-5\right) r^4 \ln r^2}{7 \left(r^2-1\right)^3}\,,\\
f_2(r) &=& \frac{2 \left(r^2+11\right) r^2}{9 \left(r^2-1\right)^2}+\frac{2 \left(5 r^2-17\right) r^4 \ln r^2}{9 \left(r^2-1\right)^3}\,,\\
%f_2(r) &=& \frac{2r_2^2 \left(r^4+10 r^2-11+2 \left(5 r^2-17\right) r^2 \ln r \right)}{9\left(r^2-1\right){}^3}\\
f_3(r) &=&\frac{2 \left(r^4+9 r^2+2\right)}{3 \left(r^2-1\right)^2}+\frac{2 \left(r^4-7 r^2-6\right) r^2 \ln r^2}{3 \left(r^2-1\right)^3}\,,\\
 f_4(r)&=&\frac{2 \left(5 r^4+25 r^2+6\right)}{7 \left(r^2-1\right)^2}+\frac{2 \left(r^4-19 r^2-18\right) r^2 \ln r^2}{7 \left(r^2-1\right)^3}\,,\\
\frac{4}{3}  f_5(r_1,r_2) &=&\frac{1+(r_1+r_2)^2- r_1^2 r_2^2}{\left(r_1^2-1\right) \left(r_2^2-1\right)}+\frac{r_1^3 \left(r_1^2+1\right)  \ln 
   r_1^2}{\left(r_1^2-1\right){}^2 \left(r_1-r_2\right)}-\frac{r_2^3 \left(r_2^2+1\right) \ln 
   r_2^2}{\left(r_1-r_2\right) \left(r_2^2-1\right){}^2}\,,\\
    \frac{7}{6}  f_6 (r_1,r_2)  &=& \frac{r_1^2+r_2^2+r_1 r_2-r_1^2 r_2^2}{\left(r_1^2-1\right)  \left(r_2^2-1\right)}+\frac{r_1^5 \ln 
r_1^2 }{\left(r_1^2-1\right){}^2  \left(r_1-r_2\right)}-\frac{r_2^5 \ln 
r_2^2}{\left(r_1-r_2\right) \left(r_2^2-1\right){}^2 }\,,\\
\frac{1}{6}f_7(r_1,r_2) &=& \frac{1+r_1 r_2}{\left(r_1^2-1\right) \left(r_2^2-1\right)}+\frac{r_1^3 \ln  r_1^2}{\left(r_1^2-1\right){}^2
   \left(r_1-r_2\right)}-\frac{r_2^3 \ln  r_2^2}{\left(r_1-r_2\right) \left(r_2^2-1\right){}^2}\,,\\
  \frac{2}{3} f_8(r_1,r_2)  &=& \frac{r_1+r_2}{\left(r_1^2-1\right) \left(r_2^2-1\right)}+\frac{r_1^4 \ln  r_1^2}{\left(r_1^2-1\right){}^2
   \left(r_1-r_2\right)}-\frac{r_2^4 \ln  r_2^2}{\left(r_1-r_2\right) \left(r_2^2-1\right){}^2} .
\end{eqnsystem}
All functions in eqs.~(\ref{sys:Fstop})--(\ref{sys:f}) are equal to 1
when they arguments approach unity, with the exception of $\wt{F}_6$
which tends to 0.

\vfill
%\newpage
\def\bf{\rm}

\small\bigskip\bigskip\footnotesize
\begin{multicols}{2}

\end{multicols}
\end{document}